\documentclass[aip,rsi,amsmath,amssymb,reprint]{revtex4-2}
\usepackage[english]{babel}
\usepackage[letterpaper,top=1.5cm,bottom=1.5cm,left=1.5cm,right=1.5cm,marginparwidth=1.5cm]{geometry}
\usepackage{dblfloatfix}
\usepackage{amsmath}
\usepackage{geometry}
\usepackage{float}
\usepackage{graphicx}
\usepackage{subfig}
\usepackage{titlesec}
\usepackage[colorlinks=true, allcolors=blue]{hyperref}
\usepackage{appendix}
\usepackage{hyperref}
\usepackage{indentfirst}
\usepackage{ragged2e}
\usepackage{color}

\begin{document}

\title{Scanning Tunneling Microscopy in high vectorial magnetic fields}

\author{Jaime Rumeu Ozores}
\affiliation{Laboratorio de Bajas Temperaturas y Altos Campos Magn\'eticos, Departamento de F\'isica de la Materia Condensada, Instituto Nicol\'as Cabrera and Condensed Matter Physics Center (IFIMAC), Unidad Asociada UAM-CSIC, Universidad Aut\'onoma de Madrid, E-28049 Madrid, Spain}

\author{Miguel \'Agueda Velasco}
\affiliation{Laboratorio de Bajas Temperaturas y Altos Campos Magn\'eticos, Departamento de F\'isica de la Materia Condensada, Instituto Nicol\'as Cabrera and Condensed Matter Physics Center (IFIMAC), Unidad Asociada UAM-CSIC, Universidad Aut\'onoma de Madrid, E-28049 Madrid, Spain}

\author{Edwin Herrera}
\affiliation{Laboratorio de Bajas Temperaturas y Altos Campos Magn\'eticos, Departamento de F\'isica de la Materia Condensada, Instituto Nicol\'as Cabrera and Condensed Matter Physics Center (IFIMAC), Unidad Asociada UAM-CSIC, Universidad Aut\'onoma de Madrid, E-28049 Madrid, Spain}

\author{Pablo Garc\'ia Talavera}
\affiliation{Laboratorio de Bajas Temperaturas y Altos Campos Magn\'eticos, Departamento de F\'isica de la Materia Condensada, Instituto Nicol\'as Cabrera and Condensed Matter Physics Center (IFIMAC), Unidad Asociada UAM-CSIC, Universidad Aut\'onoma de Madrid, E-28049 Madrid, Spain}

\author{Jose D. Berm\'udez-P\'erez}
\affiliation{Laboratorio de Bajas Temperaturas y Altos Campos Magn\'eticos, Departamento de F\'isica de la Materia Condensada, Instituto Nicol\'as Cabrera and Condensed Matter Physics Center (IFIMAC), Unidad Asociada UAM-CSIC, Universidad Aut\'onoma de Madrid, E-28049 Madrid, Spain}
\affiliation{School of Sciences and Engineering, Universidad del Rosario, Bogot\'a 111711, Colombia}

\author{Jos\'e A. Moreno}
\affiliation{Laboratorio de Bajas Temperaturas y Altos Campos Magn\'eticos, Departamento de F\'isica de la Materia Condensada, Instituto Nicol\'as Cabrera and Condensed Matter Physics Center (IFIMAC), Unidad Asociada UAM-CSIC, Universidad Aut\'onoma de Madrid, E-28049 Madrid, Spain}

\author{Paula Obladen}
\affiliation{Laboratorio de Bajas Temperaturas y Altos Campos Magn\'eticos, Departamento de F\'isica de la Materia Condensada, Instituto Nicol\'as Cabrera and Condensed Matter Physics Center (IFIMAC), Unidad Asociada UAM-CSIC, Universidad Aut\'onoma de Madrid, E-28049 Madrid, Spain}

\author{Rafael \'Alvarez Montoya}
\affiliation{ Departamento de F\'isica de la Materia Condensada, Universidad Aut\'onoma de Madrid, E-28049 Madrid, Spain}

\author{Jos\'e Navarrete}
\affiliation{SEGAINVEX, Universidad Aut\'onoma de Madrid, E-28049 Madrid, Spain}

\author{Juan Ram\'on Marijuan}
\affiliation{SEGAINVEX, Universidad Aut\'onoma de Madrid, E-28049 Madrid, Spain}

\author{Jos\'e A. Galvis}
\affiliation{School of Sciences and Engineering, Universidad del Rosario, Bogot\'a 111711, Colombia}

\author{Isabel Guillam\'on}
\affiliation{Laboratorio de Bajas Temperaturas y Altos Campos Magn\'eticos, Departamento de F\'isica de la Materia Condensada, Instituto Nicol\'as Cabrera and Condensed Matter Physics Center (IFIMAC), Unidad Asociada UAM-CSIC, Universidad Aut\'onoma de Madrid, E-28049 Madrid, Spain}

\author{Hermann Suderow}
\email[Corresponding author: ]{hermann.suderow@uam.es}
\affiliation{Laboratorio de Bajas Temperaturas y Altos Campos Magn\'eticos, Departamento de F\'isica de la Materia Condensada, Instituto Nicol\'as Cabrera and Condensed Matter Physics Center (IFIMAC), Unidad Asociada UAM-CSIC, Universidad Aut\'onoma de Madrid, E-28049 Madrid, Spain}

\begin{abstract}
{The Scanning Tunneling Microscope (STM) is a powerful instrument to study electronic density of states at surfaces down to atomic scale. Many interesting samples require studying variations as a function of the magnetic field, which is most often applied perpendicular to the surface. Conventional STM designs make it challenging to perform measurements when the magnetic field must be applied in other directions. Here we present a new STM setup installed on a rotatable platform. We have designed and built a new STM, which is small enough to allow for full rotation on a space with a diameter of 37\,mm, well below the available space within many magnets. We show that the new rotatable STM setup preserves the performance of state-of-the-art STMs in terms of noise and accuracy. Our new approach significantly enhances control over the direction of the applied magnetic field and opens exciting new possibilities to study quantum materials.}
\end{abstract}

\maketitle

\section*{Introduction}

\begin{figure}
    \centering
    \includegraphics[width=0.8\linewidth]{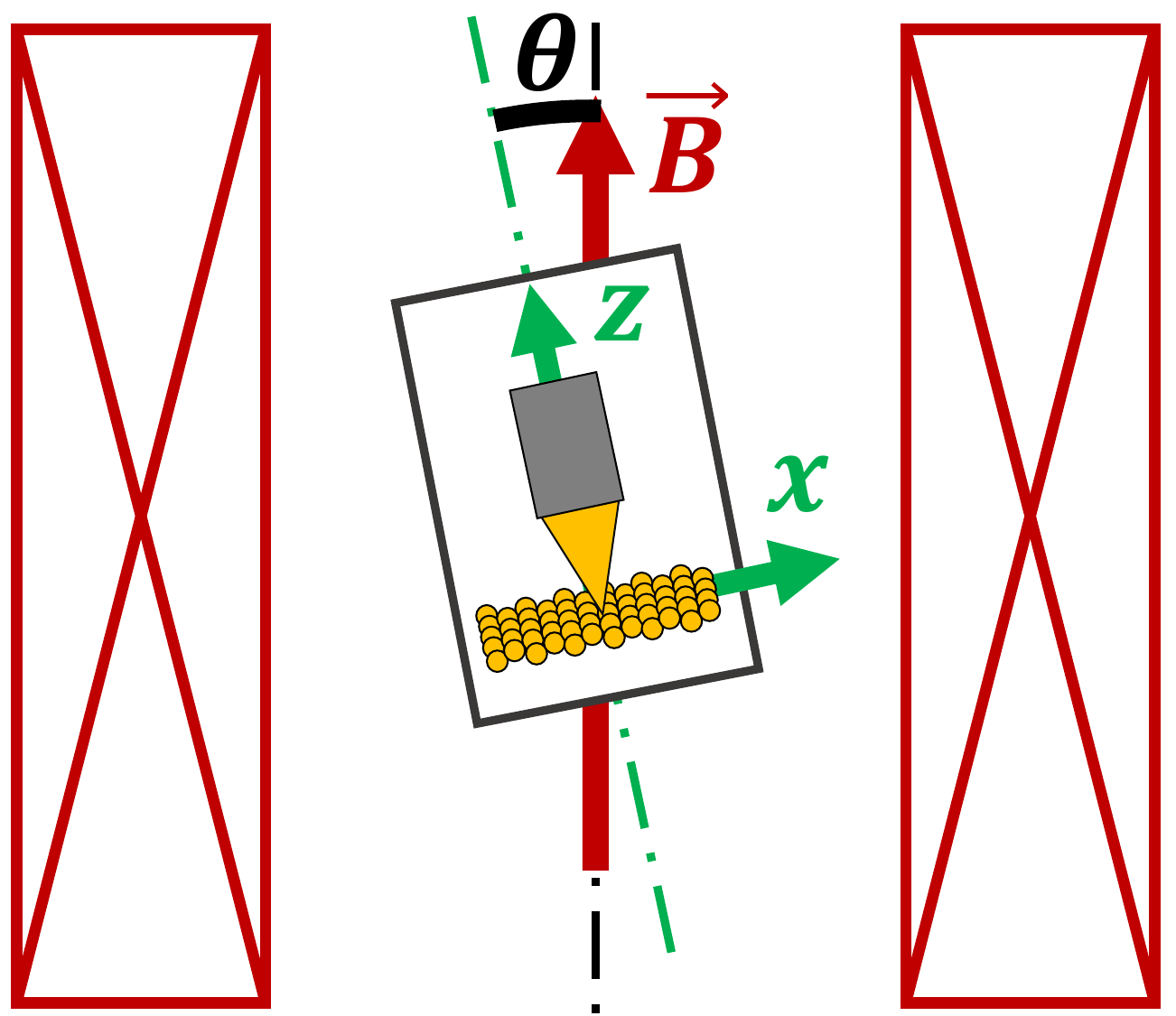}
    \caption{\justifying{We schematically show a rotatable Scanning Tunneling Microscope (STM) as a black rectangle. Within the microscope we represent schematically a scanning piezoelement (grey), a tip (orange) and a surface (orange spheres schematically representing atoms). The solenoid is shown in red. The direction of the applied magnetic field is the red arrow. The direction of the STM is varied using a rotatable platform, with the main $x$ and $z$ axis within the surface frame shown in green. The rotating angle is $\theta$.}}
    \label{Esquema}
\end{figure}

The magnetic field is a vector quantity, with magnitude and direction. Consequently, studies of a magnetic property ideally include results obtained for different magnitudes and directions of the magnetic field. For fields of several tens of mT, three-axis Helmholtz coils are widely used\,\cite{10.1063/1.1720478}. By contrast, fields in the T range require the use of superconducting solenoids. The available space often consists of a cylindrical volume with a diameter below two inches\,\cite{10.1063/5.0059394,Larbalestier2014}. Controlling the direction of the magnetic field requires either (i) combining a long solenoid along a certain direction (z) with split coils located outside the solenoid and oriented in plane (x-y), or (ii) a device that allows modifying the direction of the setup inside a long solenoid. In the first case, the split coils are located far from the measurement setup and the maximum field available in-plane is most often below 3 \,T\,\cite{PhysRevLett.69.2138, PhysRevB.50.16528,Nat.414.728, PhysRevB.66.014523, PhysRevLett.89.217003, PhysRevB.67.134501, PhysRevB.87.214504, PhysRevB.96.184502, SciRep.8.10914, ComPhy.1.30, ComPhy.2.31,10.1063/1.4905531,10.1063/1.3659412, fridmanObservationInplaneVortex2013, 10.1063/1.4995688}. The second case is the most frequently adopted solution in measurement setups which are not particularly sensitive to vibrations, including resistivity, magnetization, thermal expansion, specific heat, etc., because it allows reaching the highest magnetic fields. However, such an arrangement, schematically represented in Fig.\,\ref{Esquema}, has been comparatively less used for Scanning Tunneling Microscopes (STM)\,\cite{10.1063/1.3663611,10.1063/1.4995688,10.1063/5.0294980}. The reason is that a STM usually requires a low vibration environment and is therefore most often mounted on a rigid support without the possibility to change the orientation of the sample, typically allowing to measure only at a fixed orientation of the magnetic field\,\cite{10.1063/5.0266265, WANG2023113774,10.1063/5.0292216}. Furthermore, STMs often require the whole available space inside the solenoid, not allowing for rotation. Here we have miniaturized a STM and installed it on a rotating platform inside a superconducting solenoid. We have achieved controlled variations of the magnitude and direction of the magnetic field, without impairing the usual capabilities of cryogenic STM, i.e. atomic scale control over the position and measurement of the electronic density of states.

\begin{figure*}
    \centering
    \includegraphics[width=0.7\textwidth]{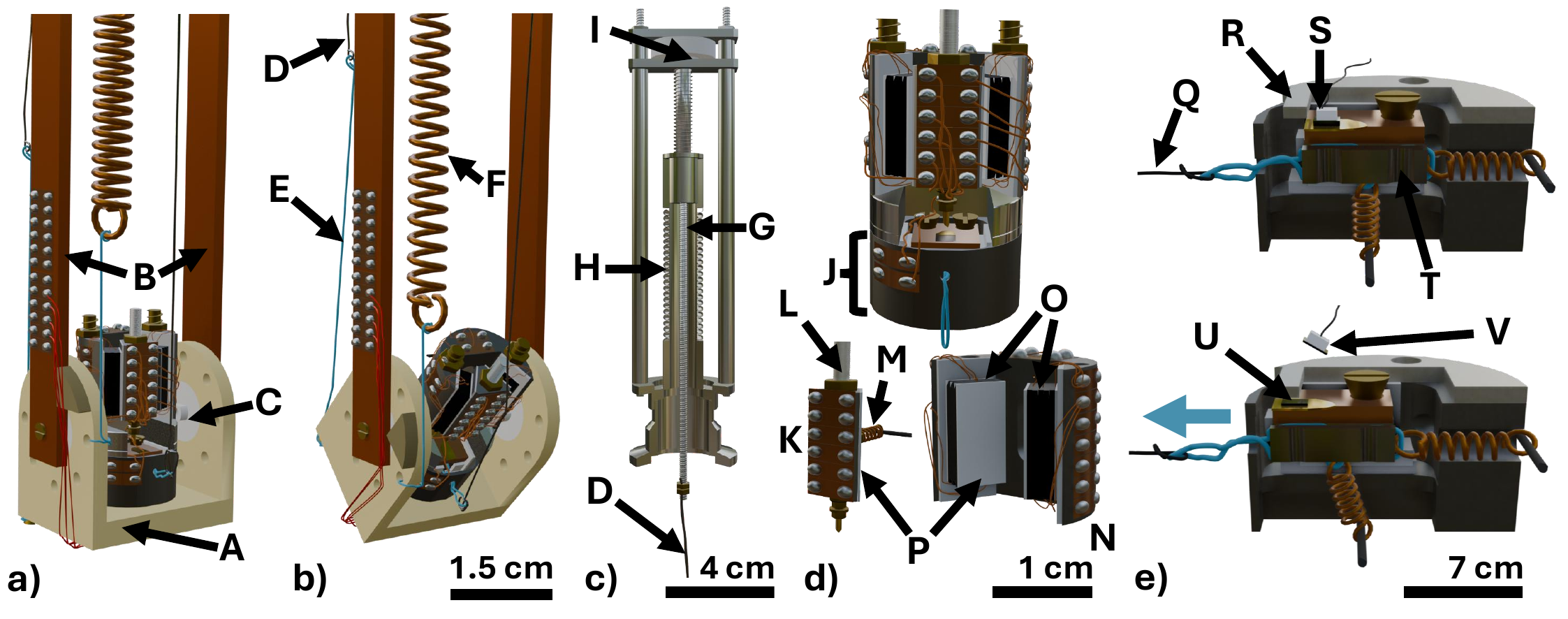}
    \caption{\justifying
{(a) Schematic of the setup, positioned such that the magnetic field is perpendicular to the surface. A is the rotatable platform. B are copper beams firmly attached to the platform. C are teflon disks which allow to firmly attach the rotatable platform to the copper beams, still allowing for in-situ rotation at cryogenic temperatures. (b) Same schematic as (a) with the platform rotated such that the magnetic field is tilted by 60° from the surface. D is a steel wire. E is a kevlar rope. F is a counteracting spring. (c) Schematic of the vacuum feedthrough actuator. G is a screwed bar. H is a bellow. I is a gearwheel. (d) Schematic drawing of the miniaturized STM. J is the STM base. K is the walker. L is the  piezotube. M is a CuBe spring used to fix the walker to the shear piezostack. N is the main head of the miniaturized STM. O are the piezo-stacks for the coarse motion. P are polished alumina. (e) Profile section of the STM base (J) and the sample showing the cleaving process. Top picture shows the sample prepared before being cleaved. Q is a steel wire attached to a second actuator. R is a beam used to cleave the sample. S is the alumina glued on the sample surface. T is the slider. Bottom picture shows the sample at the moment of being cleaved. U is the sample cleaved. V is the cleaved part of the sample.}}
    \label{Plataforma}
\end{figure*}

\section*{Description of the setup}

A relevant novel aspect of our work is the development of a rotatable platform, shown in Fig.\,\ref{Plataforma}(a,b). The rotatable platform (A in Fig.\,\ref{Plataforma}(a)) has been manufactured in polyether ether ketone (PEEK) and is attached to copper beams using two screws. The sample is located exactly at the axis of rotation and at the center of the magnet, ensuring exposure of the sample to a spatially uniform magnetic field. There is a flat supporting surface, on which we fix the miniaturized STM using screws and Kevlar ropes. Two reinforced lateral walls are screwed to a copper support. Rotation is possible thanks to two teflon rings located at either side of the assembly (C in Fig.\,\ref{Plataforma}(a)). This configuration allows continuous and dissipation-free rotation with an angular precision better than one degree. A Toshiba THS118 Hall sensor and a CCS carbon thermometer from Temati are located at the bottom of the platform to measure the direction of the magnetic field and the temperature.

The rotation is performed at cryogenic temperatures by including a long and thin stainless steel wire (0.1 mm in diameter, D in Fig.\,\ref{Plataforma}(b, c)) which we straighten using a spring and thermalize along the whole length of the cryostat by copper wires soldered to heat sinks. We can pull on the wire using a room temperature bellow actuator, shown in Fig.\,\ref{Plataforma}(c). The wire is soldered to a screw (G in Fig.\,\ref{Plataforma}(c)), which is attached to the mobile part of the bellow (H in Fig.\,\ref{Plataforma}(c)). This part is connected to a small gearwheel (I in Fig.\,\ref{Plataforma}(c)). By rotating the gearwheel in a clockwise direction we pull on the wire and rotate the platform, as schematically shown in Fig.\,\ref{Plataforma}(b). The platform is attached from the other side to a long spring (F in Fig.\,\ref{Plataforma}(b)), which reverses motion when the gearwheel rotates counterclockwise. To reduce mechanical coupling of room temperature vibrations to the platform through the steel wire, we attach the wire to a multistranded 0.18\,mm Kevlar rope (E in Fig.\,\ref{Plataforma}(b)), which is fixed to the bottom side of the rotatory platform.

The space available in the platform is considerably reduced with respect to other STM setups. In Fig.\,\ref{Plataforma}(d) we schematically show the miniaturized STM we developed to address this issue. The principle and design are similar to previously reported STM setups\,\cite{10.1063/1.3567008,10.1063/5.0059394,10.1063/5.0064511,10.1063/1.4905531,10.1063/1.3663611,10.1063/1.4995688,10.1063/5.0292216} and is based on the so-called Pan design\,\cite{10.1063/1.1149605}, with some relevant modifications. Here we have achieved a reduction of the size of the STM, both in the diameter (16 mm) and height (30 mm). The miniaturized STM essentially consists of three parts, the main head, the prism and the base. Each of them is manufactured through 3D printing in grade 3 titanium. The parts are not filled, but have a hexagonal mesh, aiming to maintain a high stiffness and reduce the weight of the miniaturized STM. The miniaturized STM head acts as a support for the walker, which hosts the piezotube. The walker is vertically moved with respect to the sample using the coarse approach system\,\cite{10.1063/5.0059394}. The coarse approach system contains two shear piezo stacks glued to the head using Stycast. The contact surfaces of both the walker and the piezo-stacks are from polished alumina covered with graphite to achieve the required combination of stick and slip, as described in Ref.\,\onlinecite{10.1063/5.0064511}. The walker is a small prism fixed to the head using a CuBe spring. The size of the walker has been reduced to the minimum needed to be able to host inside a small piezotube with external diameter of 2.5 mm. The piezotube allows for scanning 1.5 $\mu$m $\times$ 1.5 $\mu$m areas at low temperatures. The sample is located on the base on top of a movable slider, a miniaturized version of the device described in Ref.\,\onlinecite{10.1063/1.3567008}. The base is firmly fixed to the head with M2 screwed bars of brass. The required space for the whole setup is determined by the size of the miniaturized STM and the platform. The whole system requires a free diameter of 37 mm when fully rotated.

{The miniaturized STM also includes an in-situ sample and tip preparation system. In Fig.\,\ref{Plataforma}(e), we show a schematic of the sample preparation stage. We glue a 0.5 mm alumina on the sample surface (S in Fig.\,\ref{Plataforma}(e)). We cool down and we move the sample holder at cryogenic temperatures, using a system with a rope and a wire attached to a second actuator located at room temperature. The sample holder is moved below a beam, in such a way that the alumina hits the beam, thus cleaving the sample. Then, we can move the sample holder back and approach the tip to a perfectly clean surface obtained by cryogenic cleaving. To make this process compatible with the rotation stage, the actuation for sample cleaving is oriented in the opposite direction to that used to move the rotatable platform. To avoid unwanted rotation, the platform is blocked at one end when we operate the cleaving system. Within the sample holder, we also place a gold pad to condition the tip whenever needed.}

\begin{figure*}
    \includegraphics[width=0.6\linewidth]{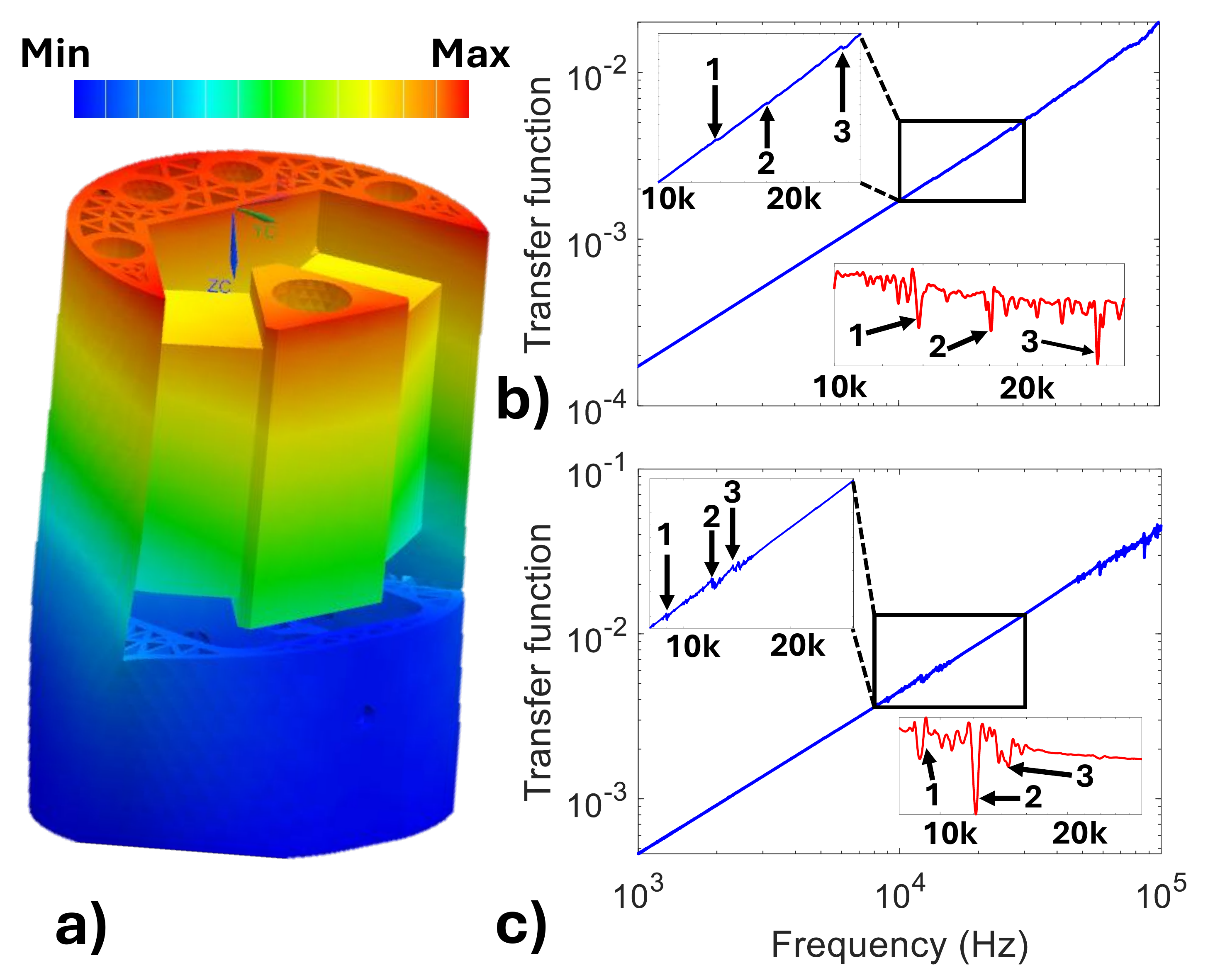}
    \caption{\justifying
   {(a) We show with a color scale the deformations of the miniaturized STM at its first resonance frequency, 11.4 kHz, obtained using finite element analysis with Siemens NX. The STM setup is drawn distorted, with an exaggerated displacement to highlight the motion corresponding to this first resonance mode. (b) Transfer function $\frac{V_{out}}{V_{in}}(\omega)$ as a function of frequency $\omega$ of the miniaturized STM obtained as described in the text. In the top left inset we show the portion of the curve inside curve the black-square of the main panel. We show with arrows and numbers the first three features observed in the transfer function. In the bottom right inset we show the derivative of the transfer function within the same frequency interval. We again show the first three resonance frequencies with arrows and numbers. These are located at 13.7 kHz, 18.3 kHz and 27.0 kHz. (c) Same as (b), but for a STM with a diameter about twice larger (see text for more details). In that case, due to the larger size of the STM, the first resonances are located at lower frequencies, 9.0 kHz, 12.1 kHz and 13.8 kHz.}}
    \label{Modos}
\end{figure*}

The vibrations influencing the performance of the miniaturized STM excite the relative motion of the tip-sample system, providing an unstable tunneling current. These vibrations can be reduced efficiently by using a STM setup which is as stiff as possible, meaning that its resonance frequency is maximized\,\cite{Voigtlander}. Reducing the mass of the STM is one possible method to increase the resonance frequency of the system. In Fig.\,\ref{Modos}(a) we show the vibration modes of the new miniaturized STM setup calculated using finite element analysis, located at about 11.4 kHz. To obtain the vibration modes of the actual setup, we applied an AC voltage through a known resistor in series with the piezo and measured the voltage at the resistor $V_{out}$, following Ref.\,\onlinecite{10.1063/1.5064442}. We sent an excitation signal $ V_{in} $ with amplitude $ V_{peak-peak}\mathrm{ = 1\,V} $ to the piezostacks providing the coarse approach in the base of the STM (N in Fig.\,\ref{Plataforma}(d)), and we swept the frequency $\omega$. A certain resonance frequency of the setup is viewed by a small feature in the transfer function $\frac{V_{out}}{V_{in}}(\omega)$. In Fig.\,\ref{Modos} (b) we show the results for the new miniaturized STM. In the bottom right inset we show the derivative of the transfer function with respect to frequency, where we can best identify the resonances. The first resonance frequency is of 13.6 kHz, very similar to the first resonance frequency obtained using finite element calculations (we show in Fig.\,\ref{Modos} (a) the calculated lowest frequency vibrational mode). In Fig.\,\ref{Modos} (c) we show the transfer function of another STM setup built previously. This STM has approximately the same shape, but a diameter of 30 mm and a height of 40 mm and was described in Ref.\,\onlinecite{10.1063/1.3567008}. The first resonance frequency here is at 9.0 kHz. Therefore, by miniaturizing the STM, we have obtained a notable increase in the resonance frequency of the STM and reduced its sensitivity to vibration.

\begin{figure*}
    \includegraphics[width=0.48\linewidth]{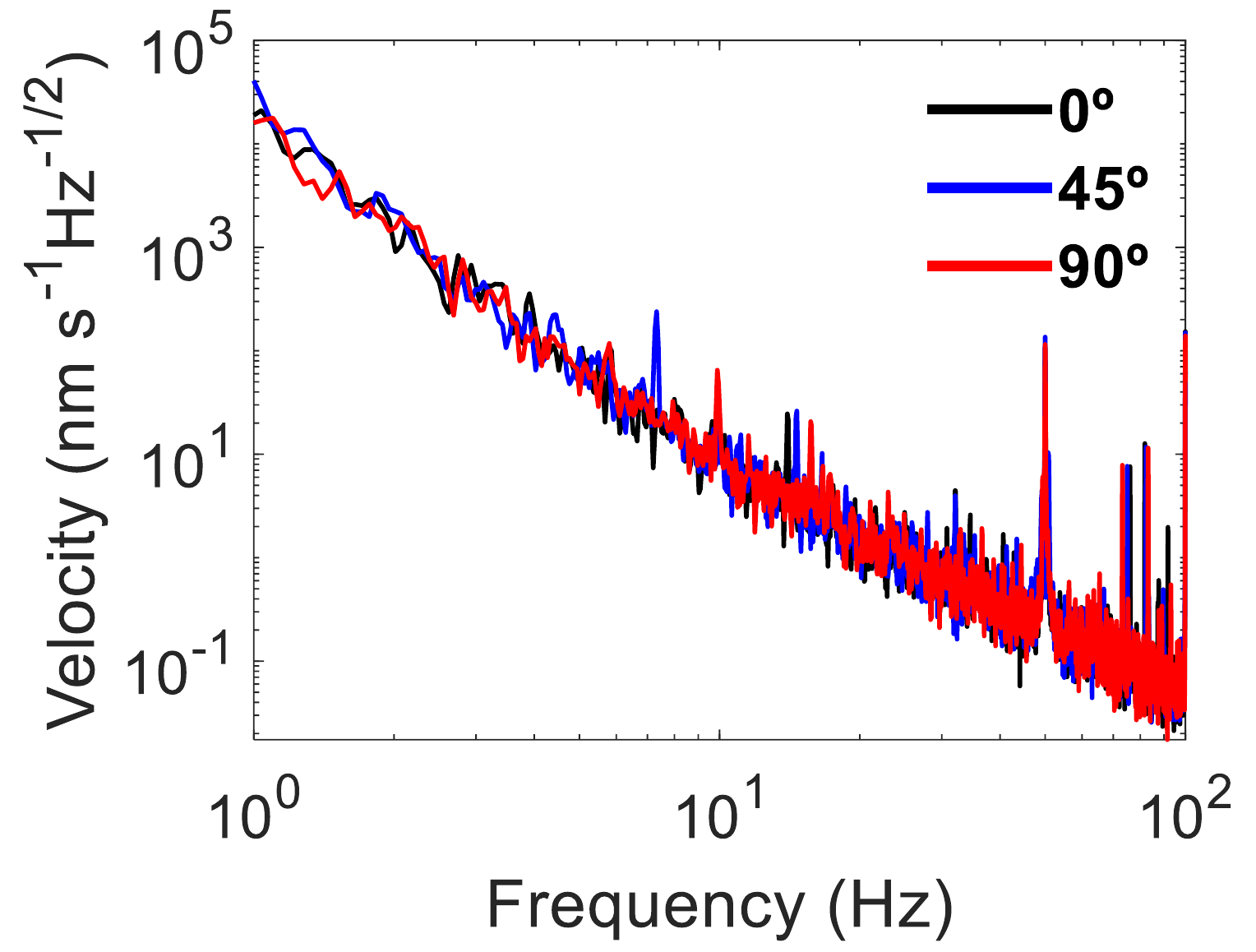}
    \caption{\justifying
   {Vibration velocity as a function of frequency with the platform rotated 0$^{\circ}$, 45$^{\circ}$ and 90$^{\circ}$. The velocity is obtained from the Fourier transform of the voltage signal measured with the accelerometer, as described in the text.}}
    \label{NOISE}
\end{figure*}

{To test the stability of the platform to low frequency vibrations (below 100 Hz) we have measured the velocity as a function of the frequency using an accelerometer (TE Connectivity) firmly attached to the platform. We used an  EGG 5113 preamplifier and then Fourier-transformed the signal to obtain the response in the frequency domain. In Fig.\,\ref{NOISE}, we show the curves obtained at 0$^{\circ}$, 45$^{\circ}$ and 90$^{\circ}$ of rotation. As usual, there is a peak at 50 Hz. However, other than that, there are no discernible features in the vibration spectrum. What is more, similar levels are obtained for all angles of rotation. The observed vibration level is also similar to the one found in previous cryogenic STM setups which use stiff and non-rotatable plaforms\,\cite{10.1063/1.5064442, 10.1063/5.0059394, 10.1063/1.4905531, 10.1063/1.5132872, 10.1063/1.4822271, 10.1063/1.3520482}. Therefore, our newly built platform allows for rotation without inducing a significant change in the level of vibrations.}

\begin{figure*}
    \centering
    \includegraphics[width=0.465\linewidth]{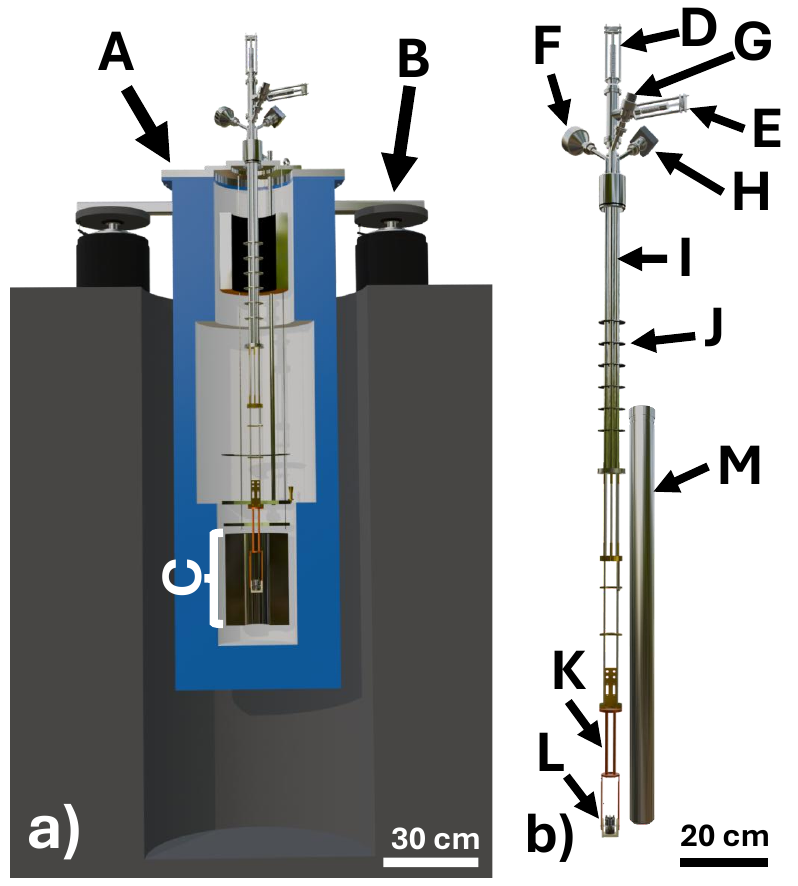}
    \caption{\justifying
    (a) Lateral view of the whole setup. The cryostat (A) is inserted in a hole in the ground (grey) and suspended on pneumatic dampers for vibration isolation (B). A 9 T superconducting magnet is located at the bottom of the cryostat (C). (b) Schematic drawing of the insert. D and E show two equivalent vacuum feedthrough actuators described more in detail in Fig.\,\ref{Plataforma} (c). F, G and H are the wire feedthroughs. I are stainless steel tubes containing the wiring. J are radiation shields. K are copper support beams holding the rotatable platform. L is the miniaturized STM located within the rotatable platform. M is the tube with an indium seal making the inner vacuum chamber. }
    \label{Criostato}
\end{figure*}

To test the platform and the miniaturized STM we used a liquid Helium cryostat from Precision Cryogenics containing a solenoid from Oxford Instruments. The platform and miniaturized STM can be, however, installed in a dilution refrigerator and in any magnet with a bore larger than 37 mm. The cryostat (A in Fig.\,\ref{Criostato}(a)) is mounted on pneumatic dampers (S-2000A pneumatic vibration isolators from Newport, B in Fig.\,\ref{Criostato}(a)). The miniaturized STM is mounted on the bottom of a long insert, shown in  Fig.\,\ref{Criostato}(b). The miniaturized STM is inside an inner vacuum chamber with Helium exchange gas for thermalization\,\cite{MONTOYA2019e00058}. The top of the insert has several vacuum feedthroughs (F-H in  Fig.\,\ref{Criostato}(b)). We apply bias voltage and measure the tunneling current through the feedthrough F in Fig.\,\ref{Criostato}(b). The feedthroughs G and H in Fig.\,\ref{Criostato}(b) are for the piezotube signals, the coarse approach system, a Hall sensor, a thermometer and a heater. Stainless steel tubes have been used to insert twisted wires for the electrical connections (I in Fig.\,\ref{Criostato}(b)). To filter unwanted high frequency noise, we add a high capacitance to ground by twisting each pair with un-insulated copper wire. The latter makes ground connections to the stainless steel tube all along the length of the wires\,\cite{10.1063/5.0059394}. Six enameled copper wires are used for the coarse approach and Hall probe, and all other wiring is of enameled manganin, with eleven wires used for the piezotube, thermometer and heater signals.

The samples investigated in this study consisted of high purity gold (99.99\,\%) obtained from Goodfellow and 2H-NbSe$_2$ grown by us. We obtained the 2H-NbSe$_2$ single crystals following the standard chemical vapor transport method using iodine\,\cite{REVOLINSKY19651029,tesissamuel}. For 2H-NbSe$_2$, first, polycrystalline powder was prerreacted by loading and mixing Nb powder (Alfa Aesar 99.99\%) and powdered Se pieces (Alfa Aesar 99.999\%) in a stoichiometric ratio into an evacuated quartz ampoule. We heated the powder to 800$^{\circ}$C for 10 days and switched the furnace off to let it cool. The resulting polycrystals were then loaded into another quartz ampoule together with 5\,mg/cm$^3$ of Iodine (Thermo Fisher Scientific 99.999\,\%) as transport agent. The source and growth zones were set to 700$^{\circ}$C/800$^{\circ}$C respectively for 48\,h. Then we applied a gradient of 750$^{\circ}$C/700$^{\circ}$C during 14 days for the growth. Shiny plates of hexagonal shape of up to 5\,mm in size were obtained.

\section*{Results}

To benchmark the performance of the system we conducted two sets of experiments at 4.2 K. First, we fabricated single atom point contacts using a tip and a sample of gold at a magnetic field of 8 T as a function of the angle between the contact and the field vector. As shown below, we achieved millions of contacts with a high precision and stability. Second, to test imaging capabilities, we have visualized the superconducting vortex lattice in 2H-NbSe$_2$ as a function of the angle between the magnetic field and the surface of the sample. With these experiments {and the characterization of vibrations shown above, we have shown that the rotatable platform allows for similar capabilities as obtained in fixed cryogenic STM platforms.}

\paragraph*{Gold atomic size contacts in tilted magnetic fields.}

\begin{figure*}[!htbp]
    \includegraphics[width=\linewidth]{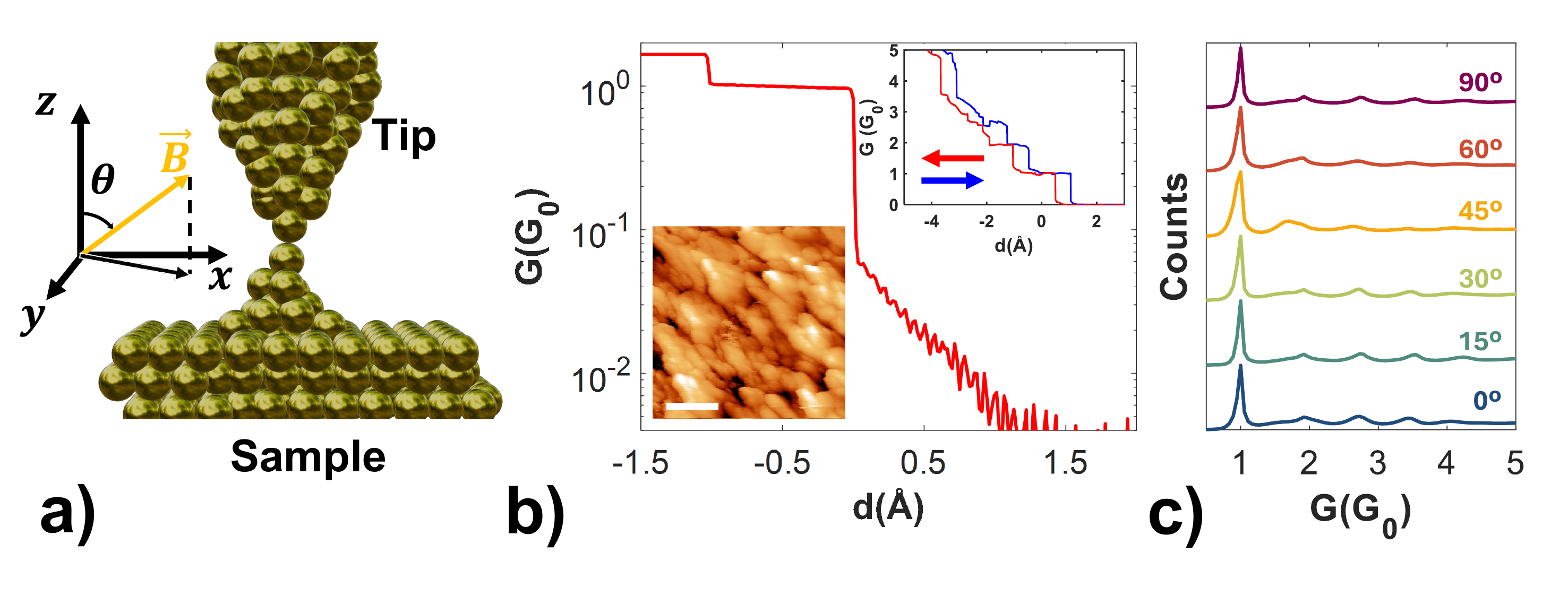}
    \caption{\justifying
    (a) Schematic drawing of a single atom point contact. Atoms are represented as light yellow spheres. Tip and sample are labeled. We show a coordinate system as black arrows and the magnetic field as an orange arrow. (b) We show as lines a few characteristic conductance $G$ vs distance $d$ curves. Red lines is for approaching tip and sample and blue line is for retracting tip from sample. The conductance is normalized to the quantum of conductance $G_0 = 2e^2/h$. The bottom left inset shows a STM topography image obtained at a tunneling current of 2\,nA and with a bias voltage of 10\,mV and at 4.2\,K. White bar is 50\,nm large. (c) We show as colored lines the conductance histograms made of about half a million of curves as the one shown in (b) for a magnetic field of 8\,T and for the angles provided in the legend.}
    \label{GOLD}
\end{figure*}

To make single atom point contacts of Au, we use tip and sample of Au, as illustrated in Fig.\,\ref{GOLD}(a). A typical experiment is shown in Fig.\,\ref{GOLD}(b). We first make a STM image, shown in the bottom left inset of Fig.\,\ref{GOLD}(b). We position the tip at the center, cut the feedback loop of the STM and approach the tip to sample. The conductance $G$ increases exponentially with decreasing distance $d$, following the dependence expected for a vacuum tunnel junction. From the slope of $log(G)$ vs $d$ (Fig.\,\ref{GOLD}(b)) we find a work function around 4 eV, as expected for Au. When the two electrodes are sufficiently close, the two Au atoms at the apex of the tip and at the surface join together through a single atom point contact\,\cite{PhysRevLett.93.116803,Agrait1993,Ohnishi1998,Yanson1998,Scheer1998,Rodrigues2000Au,Krans1995,AGRAIT2003,Sirvent1996}.  For Au, the conductance $G$ of the single atom point contact is very close to $G=G_0 = 2e^2/h$ nearly for all contacts. When further pushing the tip into the sample, the contact area becomes larger and contains several atoms. As the structure of the contact is no more well defined, the conductance is not an integer multiple of $G_0$. Nevertheless, there are characteristic features and $G(d)$ presents steps at non-integer values of $G_0$. Plotting the occurrence of a certain value of the conductance in a histogram leads to a sharp peak at $G_0$ and small peaks at non-integer multiples of $G_0$, as shown in Fig.\,\ref{GOLD}(c)\,\cite{Agrait1993,AGRAIT2003,Sirvent1996,PhysRevLett.108.205502,Sabater2018,JGRodrigo_2004,bookCuevasScheer,Requist2016,Krans1995,Ohnishi1998,Yanson1998}. We see that we obtain similar histograms for different angles (Fig.\,\ref{GOLD}(b,c). All these measurements show that the accuracy and atomic control of the STM remains the same in the new setup.

\paragraph*{Vortex lattice in tilted magnetic fields.}

\begin{figure*}[!htbp]
    \includegraphics[width=\textwidth]{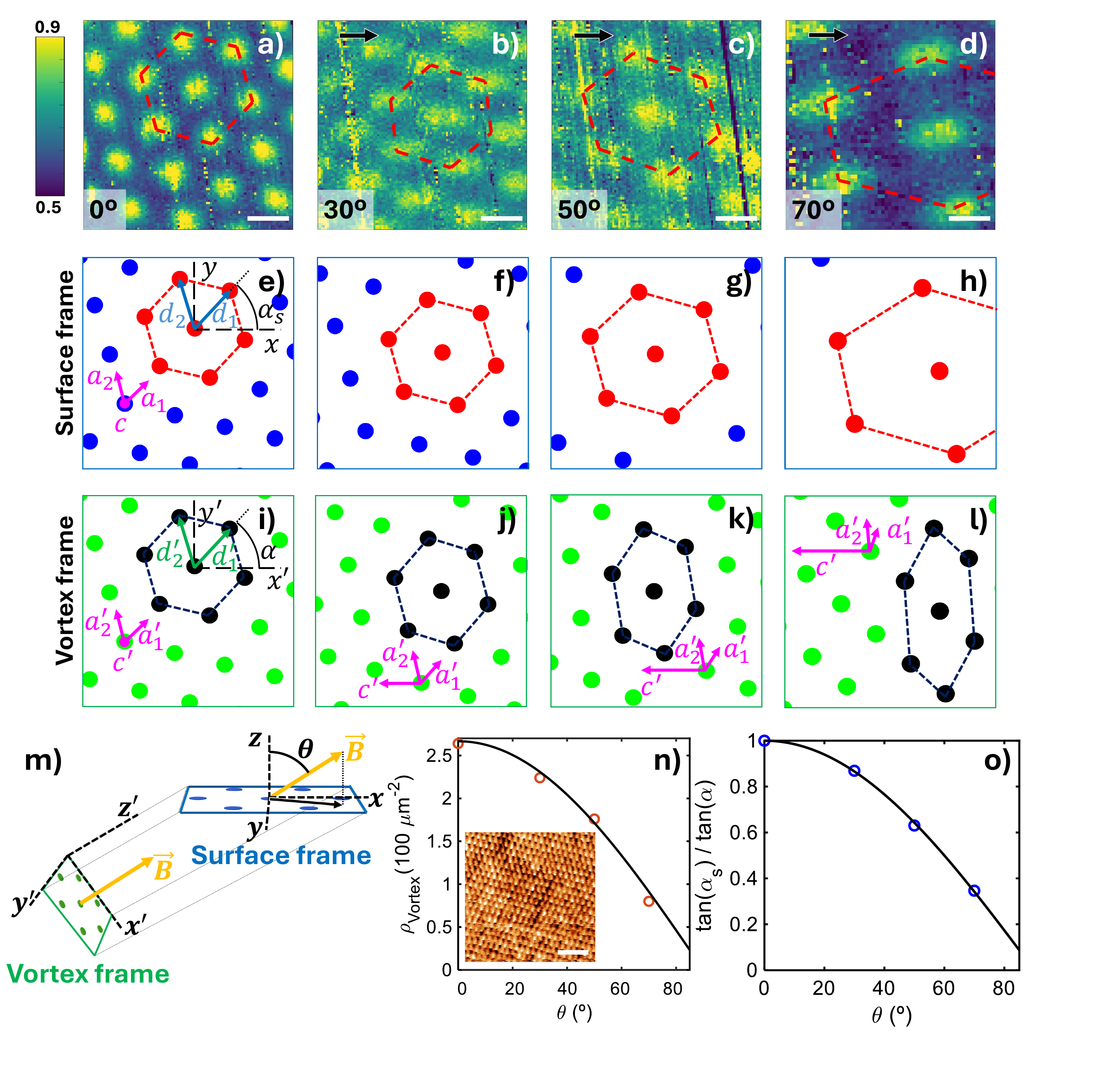}
    \caption{\justifying
    (a-d) Zero-bias tunneling conductance maps of $\mathrm{2H-NbSe_2}$ at a magnetic field of 0.5 T and at 4.2 K. The black arrows indicate the direction of the in-plane component of the magnetic field vector. White scale bar is 50 nm long. The color bar on the left  provides the normalized tunneling conductance. The tilt angle $\theta$ is indicated in each image (0$^\circ$, 30$^\circ$, 50$^\circ$ and 70$^\circ$). In (e-h) we represent as blue disks the vortex positions in the surface frame, obtained from the images shown in the top row. Red dashed line provides the hexagon formed by nearest neighbors. Crystalline axes of $\mathrm{2H-NbSe_2}$, $a_1$, $a_2$ and $c$, are shown in violet in the left panel. We show in the left panel as blue arrows $d_1$ and $d_2$, the unit cell vectors of the vortex lattice and the vortex lattice unit cell as black dashed lines. The angle $\alpha_S$ defines the orientation of the hexagonal vortex lattice with respect to the $x$-axis of the frame. In (i-l) we show the vortex lattice in the vortex lattice frame, obtained as described in the text. The vortex lattice unit cell is shown by dashed lines. Projections of crystalline axis in the vortex lattice frame $a_1'$, $a_2'$ and $c'$ are shown as violet arrows for each tilt angle. The lattice vectors of the hexagonal vortex lattice are represented as green arrows $d_1'$ and $d_2'$. (m) Schematic representation of the vortex and surface frames in a sample. Magnetic field vector is shown as an orange arrow. Its projection in the surface frame is shown by a black arrow. (n) We show as orange dots the density of vortices in the surface frame, $\rho_{vortex}(\theta)$ as a function of the tilt angle $\theta$. The inset shows a topography image with atomic resolution measured at a tunneling current of 2 nA and a bias voltage of 5 mV. White bar is 1.75 nm large. (o) We show as blue circles the ratio of the angular position of the vortex lattice in the surface and the vortex lattice frames, $\mathrm{\tan} (\alpha_S) /\mathrm{\tan} (\alpha)$ as a function of $\theta$. $\alpha_S$ ($\alpha$) is the angle between $d_1$ ($d_1'$) and the $x$-axis at the surface (vortex) frame. Continuous black lines in (n) and (o) are fits to $\mathrm{cos}(\theta)$.} 
    \label{VORTICES}
\end{figure*}

To further demonstrate the stability of the rotatable platform and the capabilities of our miniaturized STM, we have also performed measurements of the vortex lattice as a function of the orientation of the magnetic field in the layered superconductor 2H-NbSe$_2$. To visualize the superconducting vortex lattice we make maps at constant height where we acquire a full tunneling conductance vs bias voltage curve at each point (by cutting the feedback loop). {This process is particularly sensitive to vibrations, as the tip is free standing a few $ \mathrm{\AA} $ over the sample.} We then calculate the difference in the tunneling conductance at zero bias with respect to bias voltages well above the superconducting gap. Outside vortices, we observed a well developed superconducting density of states with a gap and therefore a small normalized zero bias tunneling conductance. Inside superconducting vortices we observe a high normalized zero bias tunneling conductance. Such zero bias conductance maps provide accurate images of the superconducting vortex lattice, as shown for a perpendicular field of 0.5 T in Fig.\,\ref{VORTICES}(a) (see also Refs.\,\cite{PhysRevLett.69.2138,PhysRevB.50.16528,PhysRevB.96.184502,ComPhy.1.30,ComPhy.2.31,10.1063/1.4905531,10.1063/1.3659412,fridmanObservationInplaneVortex2013}). We observe circular vortices, and the vortex lattice unit cell forms a regular hexagon whose intervortex distance $L_0\approx$ 70 nm matches with the Abrikosov lattice prediction $L_0=(4/3)^{1/4}\sqrt{\Phi_0/B}$ (where $\Phi_0$ is the flux quantum and $ B \mathrm{= 0.5\,T} $ the applied magnetic field).

{Previous studies of the vortex lattice with STM in tilted magnetic fields used mostly split superconducting coils. It was shown that the vortex lattice observed at the surface is the projection of the bulk vortex lattice to the surface and that the intervortex interaction energy should be minimized close to the surface\,\cite{PhysRevLett.69.2138,PhysRevB.50.16528,PhysRevB.96.184502,ComPhy.1.30}. In isotropic superconductors and in absence of surface effects in the intervortex interaction, the vortex lattice pattern observed at the surface is just the projection of a regular hexagon.}

{In anisotropic superconductors the superconducting coherence length might depend on the direction of the applied magnetic field\,\cite{PRB.38.4}. For example, in layered 2H-NbSe$_2$, the upper critical field perpendicular to the layers is much lower than the critical field parallel to the layers\,\cite{Dvir2018,PhysRevB.106.184514,FONER1973429,PhysRevLett.129.087002,Xi2016,Nader2014}. Therefore, the superconducting coherence length along the c-axis (perpendicular to the layers) is much smaller than the in-plane superconducting coherence length. }

In Figs.\,\ref{VORTICES}(a-d) we show the measured vortex lattice for a magnetic field of 0.5 T and several tilt angles $\theta$ with respect to the c-axis of 2H-NbSe$_2$. To analyze this quantitatively, we must distinguish between the vortex lattice frame and the surface lattice frame (see Fig.\,\ref{VORTICES}(m); see also Refs.\,\onlinecite{PRB.38.4,PhysRevLett.69.2138,PhysRevB.50.16528,PhysRevB.96.184502,ComPhy.1.30}).  By considering the anisotropy parameter, $\Gamma = (H_{c_2,ab}/H_{c_2,c})^2 $, which for  layered 2H-NbSe$_2$ is of roughly 11\,\cite{PhysRevLett.69.2138}, the vortex lattice in the vortex lattice frame (the plane perpendicular to the applied magnetic field) is a distorted hexagon, following an ellipse whose shape is related to $\Gamma$\,\cite{PhysRevLett.69.2138, PhysRevB.50.16528, PRB.38.4}. The vortex lattice in the surface frame, which is what we actually observe with the STM, is another distorted hexagon. Both can eventually be related to each other by geometry. Thus, analyzing vortex positions in the surface frame, we can deduce the shape of the vortex lattice in the vortex frame.

{We should note that for an isotropic material, the vortex lattice remains hexagonal even under tilted magnetic fields. In a uniaxial superconductor, however, the superconducting parameters become direction-dependent\,\cite{PhysRevLett.69.2138,PhysRevB.50.16528,PRB.38.4}. As we discussed above, in layered 2H-NbSe$_2$, the upper critical field perpendicular to the layers is significantly lower than that parallel to the layers. When the field is tilted, the imbalance between directions parallel and perpendicular to the plane of strong anisotropy distorts the hexagonal vortex lattice. For 2H-NbSe$_2$ we can write:

\begin{equation}
\gamma(\theta)
=
\left( 1 + \frac{1}{\Gamma}\tan^{2}\theta \right)^{1/4}
\cos^{-1/2}\theta,
\end{equation}

where $\gamma$ is the anisotropy parameter of the ellipse that encloses nearest neighbors of the distorted hexagonal vortex lattice\,\cite{Li2016, Mikitik2009, BermudezPerez2025_FeSeTiltedFields}. $1/\gamma$ increases rapidly as the direction of the magnetic field approaches the $ab$ plane. Furthermore, we can also define the orientation of the vortex lattice, $\alpha$, measured from the horizontal axes, $x'$ and the lattice basis vectors $d'_1$ and $d'_2$ in the vortex lattice frame. We can also write\,\cite{Li2016, Mikitik2009, BermudezPerez2025_FeSeTiltedFields}:

\begin{equation}
\mathbf{d}'_{1} = L_0\left( 
\gamma \cos\alpha \, \hat{\mathbf{x}}' 
+ \frac{1}{\gamma}\sin\alpha \, \hat{\mathbf{y}}' 
\right)
\end{equation}

\begin{equation}
\mathbf{d}'_{2} = L_0\left( 
\gamma \cos(\alpha+\pi/3) \, \hat{\mathbf{x}}' 
+ \frac{1}{\gamma}\sin(\alpha+\pi/3) \, \hat{\mathbf{y}}' 
\right)
\end{equation}

These vectors define the distorted vortex lattice in the vortex lattice frame and provide the basis for projecting its geometry into the surface frame.}

Let us first discuss the vortex density at the surface, i.e. the number of vortices per unit surface area. In Fig.\,\ref{VORTICES}(n) we plot the vortex density $\rho_{vortex}$ as a function of the tilt angle $\theta$, finding the expected decrease as $\mathrm{cos}(\theta)$ with increasing angle\,\cite{PhysRevB.96.184502,ComPhy.1.30}.

To further discuss the results, we note that when the field is perpendicular to the layers, the vortex lattice is oriented as shown in Fig.\,\ref{VORTICES}(e). This defines the angle $\alpha_s$ which shows the orientation of the vortex lattice for zero inclination. As discussed above, $\alpha$ is the same angle in vortex lattice frame (Fig.\,\ref{VORTICES}(i)). We plot $\mathrm{tan}(\alpha_s)/\mathrm{tan}(\alpha)$ in Fig.\,\ref{VORTICES}(o). We see that this decreases with the tilt angle as $\mathrm{tan}(\alpha_s)/\mathrm{tan}(\alpha) = \mathrm{tan}(\theta)$, as expected\,\cite{PRB.38.4,PhysRevLett.69.2138,PhysRevB.50.16528,PhysRevB.96.184502,ComPhy.1.30}.

Having resolved the positions of the vortex lattice in the vortex lattice frame we can compare its shape with respect to the crystal axis (violet arrows in Fig.\,\ref{VORTICES}(i-l)). We have obtained the directions of the Se atomic lattice vectors from atomic resolution topographies (see inset in Fig.\,\ref{VORTICES}(n)). We see that the vortex lattice is distorted. This distortion reflects the anisotropy of the critical field in $\mathrm{2H-NbSe_2}$ ($\Gamma=11$ and $H_{c2,ab} $=14.5 T as compared to $ H_{c2,c} $=4.5 T)\,\cite{PhysRevLett.69.2138,PhysRevB.50.16528, PhysB.204.1,Xi2016,Nader2014}.

\section*{Conclusions}
Our miniaturized STM with a rotatable platform shows similar characteristics as other STM systems. The miniaturized STM is smaller and has a higher resonance frequency when compared to other STM setups for studies under magnetic fields and with fields applied along different (fixed) orientations \cite{10.1063/5.0266265, MENG2020112975, 10.1063/1.4995372}. The rotatable platform provides stability to perform advanced experiments, similar to those made on setups with a fixed orientation. The rotatable miniaturized STM can be installed in cryostats reaching lower temperatures and magnets reaching higher magnetic fields, providing exciting new opportunities. It can be used to study the anisotropic properties of numerous magnetic and superconducting systems. This includes anisotropic magnetic topological semimetals and insulators, unconventional superconductors with highly anisotropic magnetic fields or superconductors with extreme forms of superconductivity for certain magnetic field directions\,\cite{NdBi,Pfleiderer,Lewin,Aoki}. For example, the heavy fermion superconductors present very anisotropic superconducting phase diagrams. Macroscopic measurements point towards a strong exchange interaction through the paramagnetic effect on Cooper pairs along certain very precise directions of the magnetic field in CeCoIn$_5$ or UPt$_3$\,\cite{Pfleiderer,Joynt02,Bisset25}. In UTe$_2$, the upper critical field is highly anisotropic providing a unique phase diagram highly sensitive to the angle of the magnetic field and a surface charge density wave whose interaction with superconductivity could be addressed by varying the direction of the magnetic field\,\cite{Lewin,Aoki,Garcia}. The interaction between superconductivity and quantum well states at surface steps of the anisotropic superconductor URu$_2$Si$_2$ can be also addressed by studying the angular dependence of the density of states on stepped surfaces\cite{Herrera}. Quantized states have been observed in two-dimensional superconductors, in which the direction of the magnetic field plays a prominent role as the critical field should be extremely high if the field is applied parallel to the superconducting plane\,\cite{Schneider,Moreno}. These and other systems require an exquisite control over the direction of the magnetic field, which can be obtained thanks to the setup described here. The new rotatable STM opens the opportunity to address relevant problems in condensed matter, which were previously more difficult to address, using microscopic measurements with atomic resolution.

\section*{Acknowledgements}

We are indebted to Frederic Hardy for suggestions on measurements at high magnetic fields and for providing the Hall probe. This work was supported by the Spanish Research State Agency (PID2023-150148OB-I00, PID2020-114071RB-I00, PDC2021-121086-I00, TED2021-130546B\-I00 and CEX2023001316-M), Comunidad de Madrid through project TEC-2024/TEC-380 “Mag4TIC” and PhD thesis support (PIPF-2023/TEC-30683, PIPF-2023/TEC-30853), the EU through grant agreement No 871106 and by the European Research Council PNICTEYES grant agreement 679080 and VECTORFIELDIMAGING grant agreement 101069239. We acknowledge the QUASURF project (SI4/PJI/2024-00199) funded by the Comunidad de Madrid through the agreement to promote and encourage research and technology transfer at the Universidad Aut\'onoma de Madrid. We acknowledge collaborations through EU program Cost CA21144 and SEGAINVEX for support in design and for construction of the miniaturized STM, the rotatable platform, the rest of the STM setup and of cryogenic components.

\section*{Appendix}
The magnetic field dependence of single atom point contacts has been studied in detail in previous work\,\cite{Strigle2015, Rodrigues2003Co, Untiedt2004, Fernandez2005, Kinikar2017, PhysRevLett.115.036601, Brun2014, PhysRevB.70.064410, PhysRevB.71.220403, SUDEROW2003264, Doudin2008, Calvo2009, Vardimon2015, Rakhmilevitch2016, PhysRevB.93.085439, Chakrabarti2022, Hayakawa2016, PhysRevB.81.134402, PhysRevB.94.144431, PhysRevB.100.214439, wu2024magnetic}. Magnetic elements provide a strong magnetoresistance. Non-magnetic elements as Au provide small features consisting of an increase in the number of contacts with resistance below $G_0$ for magnetic fields well above 10 T\,\cite{wu2024magnetic}. The electronic band structure of Au is nearly insensitive to the magnetic field. Features as Landau quantization or Zeeman splitting of bands have little or no influence in the conductance through atomic size contacts\,\cite{wu2024magnetic,PhysRevB.81.134402,Calvo2009}. Furthermore, the cyclotron radius is much larger than atomic sizes. Our results in Fig.\,\ref{GOLD} show that the atomic size contacts remain with the same shape, independently of the angle of rotation at about 8 T. It will be interesting to see if this result holds for experiments at higher magnetic fields.

%\bibliographystyle{apsrev4-2}
%\bibliography{bibliorot} 

\begin{thebibliography}{79}%
\makeatletter
\providecommand \@ifxundefined [1]{%
 \@ifx{#1\undefined}
}%
\providecommand \@ifnum [1]{%
 \ifnum #1\expandafter \@firstoftwo
 \else \expandafter \@secondoftwo
 \fi
}%
\providecommand \@ifx [1]{%
 \ifx #1\expandafter \@firstoftwo
 \else \expandafter \@secondoftwo
 \fi
}%
\providecommand \natexlab [1]{#1}%
\providecommand \enquote  [1]{``#1''}%
\providecommand \bibnamefont  [1]{#1}%
\providecommand \bibfnamefont [1]{#1}%
\providecommand \citenamefont [1]{#1}%
\providecommand \href@noop [0]{\@secondoftwo}%
\providecommand \href [0]{\begingroup \@sanitize@url \@href}%
\providecommand \@href[1]{\@@startlink{#1}\@@href}%
\providecommand \@@href[1]{\endgroup#1\@@endlink}%
\providecommand \@sanitize@url [0]{\catcode `\\12\catcode `\$12\catcode
  `\&12\catcode `\#12\catcode `\^12\catcode `\_12\catcode `\%12\relax}%
\providecommand \@@startlink[1]{}%
\providecommand \@@endlink[0]{}%
\providecommand \url  [0]{\begingroup\@sanitize@url \@url }%
\providecommand \@url [1]{\endgroup\@href {#1}{\urlprefix }}%
\providecommand \urlprefix  [0]{URL }%
\providecommand \Eprint [0]{\href }%
\providecommand \doibase [0]{https://doi.org/}%
\providecommand \selectlanguage [0]{\@gobble}%
\providecommand \bibinfo  [0]{\@secondoftwo}%
\providecommand \bibfield  [0]{\@secondoftwo}%
\providecommand \translation [1]{[#1]}%
\providecommand \BibitemOpen [0]{}%
\providecommand \bibitemStop [0]{}%
\providecommand \bibitemNoStop [0]{.\EOS\space}%
\providecommand \EOS [0]{\spacefactor3000\relax}%
\providecommand \BibitemShut  [1]{\csname bibitem#1\endcsname}%
\let\auto@bib@innerbib\@empty
%</preamble>
\bibitem [{\citenamefont {Firester}(1966)}]{10.1063/1.1720478}%
  \BibitemOpen
  \bibfield  {author} {\bibinfo {author} {\bibfnamefont {A.~H.}\ \bibnamefont
  {Firester}},\ }\bibfield  {title} {\enquote {\bibinfo {title} {Design of
  square {H}elmholtz coil systems},}\ }\href
  {https://doi.org/10.1063/1.1720478} {\bibfield  {journal} {\bibinfo
  {journal} {Review of Scientific Instruments}\ }\textbf {\bibinfo {volume}
  {37}},\ \bibinfo {pages} {1264--1265} (\bibinfo {year} {1966})}\BibitemShut
  {NoStop}%
\bibitem [{\citenamefont {Fernández-Lomana}\ \emph {et~al.}(2021)\citenamefont
  {Fernández-Lomana}, \citenamefont {Wu}, \citenamefont {Martín-Vega},
  \citenamefont {Sánchez-Barquilla}, \citenamefont {Álvarez Montoya},
  \citenamefont {Castilla}, \citenamefont {Navarrete}, \citenamefont
  {Marijuan}, \citenamefont {Herrera}, \citenamefont {Suderow},\ and\
  \citenamefont {Guillamón}}]{10.1063/5.0059394}%
  \BibitemOpen
  \bibfield  {author} {\bibinfo {author} {\bibfnamefont {M.}~\bibnamefont
  {Fernández-Lomana}}, \bibinfo {author} {\bibfnamefont {B.}~\bibnamefont
  {Wu}}, \bibinfo {author} {\bibfnamefont {F.}~\bibnamefont {Martín-Vega}},
  \bibinfo {author} {\bibfnamefont {R.}~\bibnamefont {Sánchez-Barquilla}},
  \bibinfo {author} {\bibfnamefont {R.}~\bibnamefont {Álvarez Montoya}},
  \bibinfo {author} {\bibfnamefont {J.~M.}\ \bibnamefont {Castilla}}, \bibinfo
  {author} {\bibfnamefont {J.}~\bibnamefont {Navarrete}}, \bibinfo {author}
  {\bibfnamefont {J.~R.}\ \bibnamefont {Marijuan}}, \bibinfo {author}
  {\bibfnamefont {E.}~\bibnamefont {Herrera}}, \bibinfo {author} {\bibfnamefont
  {H.}~\bibnamefont {Suderow}},\ and\ \bibinfo {author} {\bibfnamefont
  {et al.}},\ }\bibfield  {title} {\enquote {\bibinfo
  {title} {Millikelvin scanning tunneling microscope at 20/22 {T} with a
  graphite enabled stick–slip approach and an energy resolution below 8
  {$\mu$}e{V}: Application to conductance quantization at 20 {T} in single atom
  point contacts of {A}l and {A}u and to the charge density wave of
  {2H–NbSe$_2$}},}\ }\href {https://doi.org/10.1063/5.0059394} {\bibfield
  {journal} {\bibinfo  {journal} {Review of Scientific Instruments}\ }\textbf
  {\bibinfo {volume} {92}},\ \bibinfo {pages} {093701} (\bibinfo {year}
  {2021})}\BibitemShut {NoStop}%
\bibitem [{\citenamefont {Larbalestier}\ \emph {et~al.}(2014)\citenamefont
  {Larbalestier}, \citenamefont {Jiang}, \citenamefont {Trociewitz},
  \citenamefont {Kametani}, \citenamefont {Scheuerlein}, \citenamefont
  {Dalban-Canassy}, \citenamefont {Matras}, \citenamefont {Chen}, \citenamefont
  {Craig}, \citenamefont {Lee},\ and\ \citenamefont
  {Hellstrom}}]{Larbalestier2014}%
  \BibitemOpen
  \bibfield  {author} {\bibinfo {author} {\bibfnamefont {D.~C.}\ \bibnamefont
  {Larbalestier}}, \bibinfo {author} {\bibfnamefont {J.}~\bibnamefont {Jiang}},
  \bibinfo {author} {\bibfnamefont {U.~P.}\ \bibnamefont {Trociewitz}},
  \bibinfo {author} {\bibfnamefont {F.}~\bibnamefont {Kametani}}, \bibinfo
  {author} {\bibfnamefont {C.}~\bibnamefont {Scheuerlein}}, \bibinfo {author}
  {\bibfnamefont {M.}~\bibnamefont {Dalban-Canassy}}, \bibinfo {author}
  {\bibfnamefont {M.}~\bibnamefont {Matras}}, \bibinfo {author} {\bibfnamefont
  {P.}~\bibnamefont {Chen}}, \bibinfo {author} {\bibfnamefont {N.~C.}\
  \bibnamefont {Craig}}, \bibinfo {author} {\bibfnamefont {P.~J.}\ \bibnamefont
  {Lee}},\ and\ \bibinfo {author} {\bibfnamefont {et~al.}},\ }\bibfield  {title} {\enquote {\bibinfo {title} {Isotropic
  round-wire multifilament cuprate superconductor for generation of magnetic
  fields above 30 {T}},}\ }\href {https://doi.org/10.1038/nmat3887} {\bibfield
  {journal} {\bibinfo  {journal} {Nature Materials}\ }\textbf {\bibinfo
  {volume} {13}},\ \bibinfo {pages} {375--381} (\bibinfo {year}
  {2014})}\BibitemShut {NoStop}%
\bibitem [{\citenamefont {Hess}, \citenamefont {Murray},\ and\ \citenamefont
  {Waszczak}(1992)}]{PhysRevLett.69.2138}%
  \BibitemOpen
  \bibfield  {author} {\bibinfo {author} {\bibfnamefont {H.~F.}\ \bibnamefont
  {Hess}}, \bibinfo {author} {\bibfnamefont {C.~A.}\ \bibnamefont {Murray}},\
  and\ \bibinfo {author} {\bibfnamefont {J.~V.}\ \bibnamefont {Waszczak}},\
  }\bibfield  {title} {\enquote {\bibinfo {title}
  {Scanning-tunneling-microscopy study of distortion and instability of
  inclined flux-line-lattice structures in the anisotropic superconductor
  2{H}-{${\mathrm{NbSe}}_{2}$}},}\ }\href
  {https://doi.org/10.1103/PhysRevLett.69.2138} {\bibfield  {journal} {\bibinfo
   {journal} {Phys. Rev. Lett.}\ }\textbf {\bibinfo {volume} {69}},\ \bibinfo
  {pages} {2138--2141} (\bibinfo {year} {1992})}\BibitemShut {NoStop}%
\bibitem [{\citenamefont {Hess}, \citenamefont {Murray},\ and\ \citenamefont
  {Waszczak}(1994)}]{PhysRevB.50.16528}%
  \BibitemOpen
  \bibfield  {author} {\bibinfo {author} {\bibfnamefont {H.~F.}\ \bibnamefont
  {Hess}}, \bibinfo {author} {\bibfnamefont {C.~A.}\ \bibnamefont {Murray}},\
  and\ \bibinfo {author} {\bibfnamefont {J.~V.}\ \bibnamefont {Waszczak}},\
  }\bibfield  {title} {\enquote {\bibinfo {title} {Flux lattice and vortex
  structure in 2{H}-{${\mathrm{NbSe}}_{2}$} in inclined fields},}\ }\href
  {https://doi.org/10.1103/PhysRevB.50.16528} {\bibfield  {journal} {\bibinfo
  {journal} {Phys. Rev. B}\ }\textbf {\bibinfo {volume} {50}},\ \bibinfo
  {pages} {16528--16540} (\bibinfo {year} {1994})}\BibitemShut {NoStop}%
\bibitem [{\citenamefont {Grigorenko}\ \emph {et~al.}(2001)\citenamefont
  {Grigorenko}, \citenamefont {Bending}, \citenamefont {Tamegai}, \citenamefont
  {Ooi},\ and\ \citenamefont {Henini}}]{Nat.414.728}%
  \BibitemOpen
  \bibfield  {author} {\bibinfo {author} {\bibfnamefont {A.}~\bibnamefont
  {Grigorenko}}, \bibinfo {author} {\bibfnamefont {S.}~\bibnamefont {Bending}},
  \bibinfo {author} {\bibfnamefont {T.}~\bibnamefont {Tamegai}}, \bibinfo
  {author} {\bibfnamefont {S.}~\bibnamefont {Ooi}},\ and\ \bibinfo {author}
  {\bibfnamefont {M.}~\bibnamefont {Henini}},\ }\bibfield  {title} {\enquote
  {\bibinfo {title} {A one-dimensional chain state of vortex matter},}\ }\href
  {https://doi.org/10.1038/414728a} {\bibfield  {journal} {\bibinfo  {journal}
  {Nature}\ }\textbf {\bibinfo {volume} {414}},\ \bibinfo {pages} {728--731}
  (\bibinfo {year} {2001})}\BibitemShut {NoStop}%
\bibitem [{\citenamefont {Vlasko-Vlasov}\ \emph {et~al.}(2002)\citenamefont
  {Vlasko-Vlasov}, \citenamefont {Koshelev}, \citenamefont {Welp},
  \citenamefont {Crabtree},\ and\ \citenamefont
  {Kadowaki}}]{PhysRevB.66.014523}%
  \BibitemOpen
  \bibfield  {author} {\bibinfo {author} {\bibfnamefont {V.~K.}\ \bibnamefont
  {Vlasko-Vlasov}}, \bibinfo {author} {\bibfnamefont {A.}~\bibnamefont
  {Koshelev}}, \bibinfo {author} {\bibfnamefont {U.}~\bibnamefont {Welp}},
  \bibinfo {author} {\bibfnamefont {G.~W.}\ \bibnamefont {Crabtree}},\ and\
  \bibinfo {author} {\bibfnamefont {K.}~\bibnamefont {Kadowaki}},\ }\bibfield
  {title} {\enquote {\bibinfo {title} {Decoration of {J}osephson vortices by
  pancake vortices in
  {${\mathrm{Bi}}_{2}{\mathrm{Sr}}_{2}{\mathrm{CaCu}}_{2}{\mathrm{O}}_{8+d}$}},}\
  }\href {https://doi.org/10.1103/PhysRevB.66.014523} {\bibfield  {journal}
  {\bibinfo  {journal} {Phys. Rev. B}\ }\textbf {\bibinfo {volume} {66}},\
  \bibinfo {pages} {014523} (\bibinfo {year} {2002})}\BibitemShut {NoStop}%
\bibitem [{\citenamefont {Grigorenko}\ \emph {et~al.}(2002)\citenamefont
  {Grigorenko}, \citenamefont {Bending}, \citenamefont {Koshelev},
  \citenamefont {Clem}, \citenamefont {Tamegai},\ and\ \citenamefont
  {Ooi}}]{PhysRevLett.89.217003}%
  \BibitemOpen
  \bibfield  {author} {\bibinfo {author} {\bibfnamefont {A.~N.}\ \bibnamefont
  {Grigorenko}}, \bibinfo {author} {\bibfnamefont {S.~J.}\ \bibnamefont
  {Bending}}, \bibinfo {author} {\bibfnamefont {A.~E.}\ \bibnamefont
  {Koshelev}}, \bibinfo {author} {\bibfnamefont {J.~R.}\ \bibnamefont {Clem}},
  \bibinfo {author} {\bibfnamefont {T.}~\bibnamefont {Tamegai}},\ and\ \bibinfo
  {author} {\bibfnamefont {S.}~\bibnamefont {Ooi}},\ }\bibfield  {title}
  {\enquote {\bibinfo {title} {Visualization of interacting crossing vortex
  lattices in the presence of quenched disorder},}\ }\href
  {https://doi.org/10.1103/PhysRevLett.89.217003} {\bibfield  {journal}
  {\bibinfo  {journal} {Phys. Rev. Lett.}\ }\textbf {\bibinfo {volume} {89}},\
  \bibinfo {pages} {217003} (\bibinfo {year} {2002})}\BibitemShut {NoStop}%
\bibitem [{\citenamefont {Tokunaga}\ \emph {et~al.}(2003)\citenamefont
  {Tokunaga}, \citenamefont {Tamegai}, \citenamefont {Fasano},\ and\
  \citenamefont {de~la Cruz}}]{PhysRevB.67.134501}%
  \BibitemOpen
  \bibfield  {author} {\bibinfo {author} {\bibfnamefont {M.}~\bibnamefont
  {Tokunaga}}, \bibinfo {author} {\bibfnamefont {T.}~\bibnamefont {Tamegai}},
  \bibinfo {author} {\bibfnamefont {Y.}~\bibnamefont {Fasano}},\ and\ \bibinfo
  {author} {\bibfnamefont {F.}~\bibnamefont {de~la Cruz}},\ }\bibfield  {title}
  {\enquote {\bibinfo {title} {Direct observations of the vortex chain state in
  {${\mathrm{Bi}}_{2}{\mathrm{Sr}}_{2}{\mathrm{CaCu}}_{2}{\mathrm{O}}_{8+y}$}
  by {B}itter decoration},}\ }\href
  {https://doi.org/10.1103/PhysRevB.67.134501} {\bibfield  {journal} {\bibinfo
  {journal} {Phys. Rev. B}\ }\textbf {\bibinfo {volume} {67}},\ \bibinfo
  {pages} {134501} (\bibinfo {year} {2003})}\BibitemShut {NoStop}%
\bibitem [{\citenamefont {Galvis}\ \emph {et~al.}(2013)\citenamefont {Galvis},
  \citenamefont {Suderow}, \citenamefont {Vieira}, \citenamefont {Bud'ko},\
  and\ \citenamefont {Canfield}}]{PhysRevB.87.214504}%
  \BibitemOpen
  \bibfield  {author} {\bibinfo {author} {\bibfnamefont {J.~A.}\ \bibnamefont
  {Galvis}}, \bibinfo {author} {\bibfnamefont {H.}~\bibnamefont {Suderow}},
  \bibinfo {author} {\bibfnamefont {S.}~\bibnamefont {Vieira}}, \bibinfo
  {author} {\bibfnamefont {S.~L.}\ \bibnamefont {Bud'ko}},\ and\ \bibinfo
  {author} {\bibfnamefont {P.~C.}\ \bibnamefont {Canfield}},\ }\bibfield
  {title} {\enquote {\bibinfo {title} {Scanning tunneling microscopy in the
  superconductor {LaSb${}_{2}$}},}\ }\href
  {https://doi.org/10.1103/PhysRevB.87.214504} {\bibfield  {journal} {\bibinfo
  {journal} {Phys. Rev. B}\ }\textbf {\bibinfo {volume} {87}},\ \bibinfo
  {pages} {214504} (\bibinfo {year} {2013})}\BibitemShut {NoStop}%
\bibitem [{\citenamefont {Herrera}\ \emph {et~al.}(2017)\citenamefont
  {Herrera}, \citenamefont {Guillam\'on}, \citenamefont {Galvis}, \citenamefont
  {Correa}, \citenamefont {Fente}, \citenamefont {Vieira}, \citenamefont
  {Suderow}, \citenamefont {Martynovich},\ and\ \citenamefont
  {Kogan}}]{PhysRevB.96.184502}%
  \BibitemOpen
  \bibfield  {author} {\bibinfo {author} {\bibfnamefont {E.}~\bibnamefont
  {Herrera}}, \bibinfo {author} {\bibfnamefont {I.}~\bibnamefont
  {Guillam\'on}}, \bibinfo {author} {\bibfnamefont {J.~A.}\ \bibnamefont
  {Galvis}}, \bibinfo {author} {\bibfnamefont {A.}~\bibnamefont {Correa}},
  \bibinfo {author} {\bibfnamefont {A.}~\bibnamefont {Fente}}, \bibinfo
  {author} {\bibfnamefont {S.}~\bibnamefont {Vieira}}, \bibinfo {author}
  {\bibfnamefont {H.}~\bibnamefont {Suderow}}, \bibinfo {author} {\bibfnamefont
  {A.~Y.}\ \bibnamefont {Martynovich}},\ and\ \bibinfo {author} {\bibfnamefont
  {V.~G.}\ \bibnamefont {Kogan}},\ }\bibfield  {title} {\enquote {\bibinfo
  {title} {Subsurface bending and reorientation of tilted vortex lattices in
  bulk isotropic superconductors due to {C}oulomb-like repulsion at the
  surface},}\ }\href {https://doi.org/10.1103/PhysRevB.96.184502} {\bibfield
  {journal} {\bibinfo  {journal} {Phys. Rev. B}\ }\textbf {\bibinfo {volume}
  {96}},\ \bibinfo {pages} {184502} (\bibinfo {year} {2017})}\BibitemShut
  {NoStop}%
\bibitem [{\citenamefont {Curran}\ \emph {et~al.}(2018)\citenamefont {Curran},
  \citenamefont {Mohammed}, \citenamefont {Bending}, \citenamefont {Koshelev},
  \citenamefont {Tsuchiya},\ and\ \citenamefont {Tamegai}}]{SciRep.8.10914}%
  \BibitemOpen
  \bibfield  {author} {\bibinfo {author} {\bibfnamefont {P.~J.}\ \bibnamefont
  {Curran}}, \bibinfo {author} {\bibfnamefont {H.~A.}\ \bibnamefont
  {Mohammed}}, \bibinfo {author} {\bibfnamefont {S.~J.}\ \bibnamefont
  {Bending}}, \bibinfo {author} {\bibfnamefont {A.~E.}\ \bibnamefont
  {Koshelev}}, \bibinfo {author} {\bibfnamefont {Y.}~\bibnamefont {Tsuchiya}},\
  and\ \bibinfo {author} {\bibfnamefont {T.}~\bibnamefont {Tamegai}},\
  }\bibfield  {title} {\enquote {\bibinfo {title} {Tuning the structure of the
  {J}osephson vortex lattice in {Bi${}_{2}$Sr${}_{2}$CaCu${}_{2}$O${}_{8 +
  \delta}$} single crystals with pancake vortices},}\ }\href
  {https://doi.org/10.1038/s41598-018-28681-7} {\bibfield  {journal} {\bibinfo
  {journal} {Scientific Reports}\ }\textbf {\bibinfo {volume} {8}},\ \bibinfo
  {pages} {10914} (\bibinfo {year} {2018})}\BibitemShut {NoStop}%
\bibitem [{\citenamefont {Galvis}\ \emph {et~al.}(2018)\citenamefont {Galvis},
  \citenamefont {Herrera}, \citenamefont {Berthod}, \citenamefont {Vieira},
  \citenamefont {Guillamón},\ and\ \citenamefont {Suderow}}]{ComPhy.1.30}%
  \BibitemOpen
  \bibfield  {author} {\bibinfo {author} {\bibfnamefont {J.~A.}\ \bibnamefont
  {Galvis}}, \bibinfo {author} {\bibfnamefont {E.}~\bibnamefont {Herrera}},
  \bibinfo {author} {\bibfnamefont {C.}~\bibnamefont {Berthod}}, \bibinfo
  {author} {\bibfnamefont {S.}~\bibnamefont {Vieira}}, \bibinfo {author}
  {\bibfnamefont {I.}~\bibnamefont {Guillamón}},\ and\ \bibinfo {author}
  {\bibfnamefont {H.}~\bibnamefont {Suderow}},\ }\bibfield  {title} {\enquote
  {\bibinfo {title} {Tilted vortex cores and superconducting gap anisotropy in
  2{H}-{${\mathrm{NbSe}}_{2}$}},}\ }\href
  {https://doi.org/10.1038/s42005-018-0028-1} {\bibfield  {journal} {\bibinfo
  {journal} {Communications Physics}\ }\textbf {\bibinfo {volume} {1}},\
  \bibinfo {pages} {30} (\bibinfo {year} {2018})}\BibitemShut {NoStop}%
\bibitem [{\citenamefont {Correa}\ \emph {et~al.}(2019)\citenamefont {Correa},
  \citenamefont {Mompeán}, \citenamefont {Guillamón}, \citenamefont
  {Herrera}, \citenamefont {García-Hernández}, \citenamefont {Yamamoto},
  \citenamefont {Kashiwagi}, \citenamefont {Kadowaki}, \citenamefont {Buzdin},
  \citenamefont {Suderow},\ and\ \citenamefont {Munuera}}]{ComPhy.2.31}%
  \BibitemOpen
  \bibfield  {author} {\bibinfo {author} {\bibfnamefont {A.}~\bibnamefont
  {Correa}}, \bibinfo {author} {\bibfnamefont {F.}~\bibnamefont {Mompeán}},
  \bibinfo {author} {\bibfnamefont {I.}~\bibnamefont {Guillamón}}, \bibinfo
  {author} {\bibfnamefont {E.}~\bibnamefont {Herrera}}, \bibinfo {author}
  {\bibfnamefont {M.}~\bibnamefont {García-Hernández}}, \bibinfo {author}
  {\bibfnamefont {T.}~\bibnamefont {Yamamoto}}, \bibinfo {author}
  {\bibfnamefont {T.}~\bibnamefont {Kashiwagi}}, \bibinfo {author}
  {\bibfnamefont {K.}~\bibnamefont {Kadowaki}}, \bibinfo {author}
  {\bibfnamefont {A.~I.}\ \bibnamefont {Buzdin}}, \bibinfo {author}
  {\bibfnamefont {H.}~\bibnamefont {Suderow}},\ and\ \bibinfo {author}
  {\bibfnamefont {et~al.}},\ }\bibfield  {title} {\enquote
  {\bibinfo {title} {Attractive interaction between superconducting vortices in
  tilted magnetic fields},}\ }\href {https://doi.org/10.1038/s42005-019-0132-x}
  {\bibfield  {journal} {\bibinfo  {journal} {Communications Physics}\ }\textbf
  {\bibinfo {volume} {2}},\ \bibinfo {pages} {31} (\bibinfo {year}
  {2019})}\BibitemShut {NoStop}%
\bibitem [{\citenamefont {Galvis}\ \emph {et~al.}(2015)\citenamefont {Galvis},
  \citenamefont {Herrera}, \citenamefont {Guillamón}, \citenamefont
  {Azpeitia}, \citenamefont {Luccas}, \citenamefont {Munuera}, \citenamefont
  {Cuenca}, \citenamefont {Higuera}, \citenamefont {Díaz}, \citenamefont
  {Pazos}, \citenamefont {García-Hernandez}, \citenamefont {Buendía},
  \citenamefont {Vieira},\ and\ \citenamefont {Suderow}}]{10.1063/1.4905531}%
  \BibitemOpen
  \bibfield  {author} {\bibinfo {author} {\bibfnamefont {J.~A.}\ \bibnamefont
  {Galvis}}, \bibinfo {author} {\bibfnamefont {E.}~\bibnamefont {Herrera}},
  \bibinfo {author} {\bibfnamefont {I.}~\bibnamefont {Guillamón}}, \bibinfo
  {author} {\bibfnamefont {J.}~\bibnamefont {Azpeitia}}, \bibinfo {author}
  {\bibfnamefont {R.~F.}\ \bibnamefont {Luccas}}, \bibinfo {author}
  {\bibfnamefont {C.}~\bibnamefont {Munuera}}, \bibinfo {author} {\bibfnamefont
  {M.}~\bibnamefont {Cuenca}}, \bibinfo {author} {\bibfnamefont {J.~A.}\
  \bibnamefont {Higuera}}, \bibinfo {author} {\bibfnamefont {N.}~\bibnamefont
  {Díaz}}, \bibinfo {author} {\bibfnamefont {M.}~\bibnamefont {Pazos}},
  \bibinfo {author} {\bibfnamefont {et~al.}}}\bibfield  {title} {\enquote
  {\bibinfo {title} {Three axis vector magnet set-up for cryogenic scanning
  probe microscopy},}\ }\href {https://doi.org/10.1063/1.4905531} {\bibfield
  {journal} {\bibinfo  {journal} {Review of Scientific Instruments}\ }\textbf
  {\bibinfo {volume} {86}},\ \bibinfo {pages} {013706} (\bibinfo {year}
  {2015})}\BibitemShut {NoStop}%
\bibitem [{\citenamefont {Fridman}\ \emph {et~al.}(2011)\citenamefont
  {Fridman}, \citenamefont {Kloc}, \citenamefont {Petrovic},\ and\
  \citenamefont {Wei}}]{10.1063/1.3659412}%
  \BibitemOpen
  \bibfield  {author} {\bibinfo {author} {\bibfnamefont {I.}~\bibnamefont
  {Fridman}}, \bibinfo {author} {\bibfnamefont {C.}~\bibnamefont {Kloc}},
  \bibinfo {author} {\bibfnamefont {C.}~\bibnamefont {Petrovic}},\ and\
  \bibinfo {author} {\bibfnamefont {J.~Y.~T.}\ \bibnamefont {Wei}},\ }\bibfield
   {title} {\enquote {\bibinfo {title} {Lateral imaging of the superconducting
  vortex lattice using {D}oppler-modulated scanning tunneling microscopy},}\
  }\href {https://doi.org/10.1063/1.3659412} {\bibfield  {journal} {\bibinfo
  {journal} {Applied Physics Letters}\ }\textbf {\bibinfo {volume} {99}},\
  \bibinfo {pages} {192505} (\bibinfo {year} {2011})}\BibitemShut {NoStop}%
\bibitem [{\citenamefont {Fridman}\ \emph {et~al.}(2013)\citenamefont
  {Fridman}, \citenamefont {Kloc}, \citenamefont {Petrovic},\ and\
  \citenamefont {Wei}}]{fridmanObservationInplaneVortex2013}%
  \BibitemOpen
  \bibfield  {author} {\bibinfo {author} {\bibfnamefont {I.}~\bibnamefont
  {Fridman}}, \bibinfo {author} {\bibfnamefont {C.}~\bibnamefont {Kloc}},
  \bibinfo {author} {\bibfnamefont {C.}~\bibnamefont {Petrovic}},\ and\
  \bibinfo {author} {\bibfnamefont {J.~Y.~T.}\ \bibnamefont {Wei}},\
  }\href@noop {} {\enquote {\bibinfo {title} {Observation of an in-plane vortex
  lattice transition in the multiband superconductor {{2H-NbSe2}} using
  scanning tunneling spectroscopy},}\ } (\bibinfo {year} {2013}),\ \Eprint
  {https://arxiv.org/abs/1303.3559} {arXiv:1303.3559 [cond-mat]} \BibitemShut
  {NoStop}%
\bibitem [{\citenamefont {Trainer}\ \emph {et~al.}(2017)\citenamefont
  {Trainer}, \citenamefont {Yim}, \citenamefont {McLaren},\ and\ \citenamefont
  {Wahl}}]{10.1063/1.4995688}%
  \BibitemOpen
  \bibfield  {author} {\bibinfo {author} {\bibfnamefont {C.}~\bibnamefont
  {Trainer}}, \bibinfo {author} {\bibfnamefont {C.~M.}\ \bibnamefont {Yim}},
  \bibinfo {author} {\bibfnamefont {M.}~\bibnamefont {McLaren}},\ and\ \bibinfo
  {author} {\bibfnamefont {P.}~\bibnamefont {Wahl}},\ }\bibfield  {title}
  {\enquote {\bibinfo {title} {{Cryogenic STM in 3D vector magnetic fields
  realized through a rotatable insert}},}\ }\href
  {https://doi.org/10.1063/1.4995688} {\bibfield  {journal} {\bibinfo
  {journal} {Review of Scientific Instruments}\ }\textbf {\bibinfo {volume}
  {88}},\ \bibinfo {pages} {093705} (\bibinfo {year} {2017})}\BibitemShut
  {NoStop}%
\bibitem {10.1063/1.3663611}%
  \BibitemOpen
  \bibfield  {author} {\bibinfo {author} {\bibfnamefont {S.~C.}~\bibnamefont
  {White}}, \bibinfo {author} {\bibfnamefont {U.~R.}\ \bibnamefont {Singh}},
  and\ \bibinfo
  {author} {\bibfnamefont {P.}~\bibnamefont {Wahl}},\ }\bibfield  {title}
  {\enquote {\bibinfo {title} {{A stiff scanning tunneling microscopy head for measurement at low temperatures and in high magnetic fields}},}\ }\href
  {https://doi.org/10.1063/1.3663611} {\bibfield  {journal} {\bibinfo
  {journal} {Review of Scientific Instruments}\ }\textbf {\bibinfo {volume}
  {82}},\ \bibinfo {pages} {113708 } (\bibinfo {year} {2011})}\BibitemShut
  {NoStop}%
\bibitem {10.1063/5.0294980}%
  \BibitemOpen
  \bibfield  {author} {\bibinfo {author} {\bibfnamefont {Yue}~\bibnamefont
  {Gao}}, \bibinfo {author} {\bibfnamefont {Wenjie}\ \bibnamefont {Meng}},
  \bibinfo {author} {\bibfnamefont {Yuchen}\ \bibnamefont {Zhang}},
  \bibinfo {author} {\bibfnamefont {Shuai}\ \bibnamefont {Dong}},
  \bibinfo {author} {\bibfnamefont {Aile}\ \bibnamefont {Wang}},
  \bibinfo {author} {\bibfnamefont {Min}\ \bibnamefont {Zhang}},
  \bibinfo {author} {\bibfnamefont {Jihao}\ \bibnamefont {Wang}},
  \bibinfo {author} {\bibfnamefont {Yubin}\ \bibnamefont {Hou}},
  and\ \bibinfo
  {author} {\bibfnamefont {Qingyou}~\bibnamefont {Lu}},\ }
  \bibfield  {title} {\enquote {\bibinfo {title} {{Compact piezo-driven rotatable magnetic force microscope in a cryogen-free magnet}},}\ }\href
  {https://doi.org/10.1063/5.0294980} {\bibfield  {journal} {\bibinfo
  {journal} {Review of Scientific Instruments}\ }\textbf {\bibinfo {volume}
  {96}},\ \bibinfo {pages} {123707} (\bibinfo {year} {2025})}\BibitemShut
  {NoStop}%
\bibitem {10.1063/5.0292216}%
  \BibitemOpen
  \bibfield  {author} {\bibinfo {author} {\bibfnamefont {Olivia}\ \bibnamefont {Armitage}},
  \bibinfo {author} {\bibfnamefont {Haibiao}\ \bibnamefont {Zhou}},
  \bibinfo {author} {\bibfnamefont {Bruno}\ \bibnamefont {Saika}},
  \bibinfo {author} {\bibfnamefont {Daniel A.}\ \bibnamefont {Vajner}},
  \bibinfo {author} {\bibfnamefont {Martin}\ \bibnamefont {McLaren}},
  and\ \bibinfo
  {author} {\bibfnamefont {Peter}~\bibnamefont {Wahl}},\ }
    \bibfield  {title}
  {\enquote {\bibinfo {title} {{Design for a 1 K pot for a low-temperature ultra-high vacuum scanning tunneling microscope}},}\ }\href
  {https://doi.org/10.1063/5.0292216} {\bibfield  {journal} {\bibinfo
  {journal} {Review of Scientific Instruments}\ }\textbf {\bibinfo {volume}
  {96}},\ \bibinfo {pages} {123906} (\bibinfo {year} {2025})}\BibitemShut
  {NoStop}%
\bibitem [{\citenamefont {Wu}\ \emph {et~al.}(2025)\citenamefont {Wu},
  \citenamefont {Wang}, \citenamefont {Dong}, \citenamefont {Li}, \citenamefont
  {Liang}, \citenamefont {Wang}, \citenamefont {Zhang}, \citenamefont {Zhang},
  \citenamefont {Feng}, \citenamefont {Meng}, \citenamefont {Hou},\ and\
  \citenamefont {Lu}}]{10.1063/5.0266265}%
  \BibitemOpen
  \bibfield  {author} {\bibinfo {author} {\bibfnamefont {D.}~\bibnamefont
  {Wu}}, \bibinfo {author} {\bibfnamefont {J.}~\bibnamefont {Wang}}, \bibinfo
  {author} {\bibfnamefont {S.}~\bibnamefont {Dong}}, \bibinfo {author}
  {\bibfnamefont {Z.}~\bibnamefont {Li}}, \bibinfo {author} {\bibfnamefont
  {R.}~\bibnamefont {Liang}}, \bibinfo {author} {\bibfnamefont
  {A.}~\bibnamefont {Wang}}, \bibinfo {author} {\bibfnamefont {M.}~\bibnamefont
  {Zhang}}, \bibinfo {author} {\bibfnamefont {J.}~\bibnamefont {Zhang}},
  \bibinfo {author} {\bibfnamefont {Q.}~\bibnamefont {Feng}}, \bibinfo {author}
  {\bibfnamefont {W.}~\bibnamefont {Meng}}, \bibinfo {author} {\bibfnamefont
  {et~al.~}}}\bibfield  {title} {\enquote {\bibinfo {title}
  {{35 T water-cooled magnet scanning tunneling microscope for in-plane
  magnetic field measurement}},}\ }\href {https://doi.org/10.1063/5.0266265}
  {\bibfield  {journal} {\bibinfo  {journal} {Review of Scientific
  Instruments}\ }\textbf {\bibinfo {volume} {96}},\ \bibinfo {pages} {083708}
  (\bibinfo {year} {2025})}\BibitemShut {NoStop}%
\bibitem [{\citenamefont {Wang}\ \emph {et~al.}(2023)\citenamefont {Wang},
  \citenamefont {Li}, \citenamefont {Meng}, \citenamefont {Hou}, \citenamefont
  {Lu},\ and\ \citenamefont {Lu}}]{WANG2023113774}%
  \BibitemOpen
  \bibfield  {author} {\bibinfo {author} {\bibfnamefont {J.}~\bibnamefont
  {Wang}}, \bibinfo {author} {\bibfnamefont {W.}~\bibnamefont {Li}}, \bibinfo
  {author} {\bibfnamefont {W.}~\bibnamefont {Meng}}, \bibinfo {author}
  {\bibfnamefont {Y.}~\bibnamefont {Hou}}, \bibinfo {author} {\bibfnamefont
  {Y.}~\bibnamefont {Lu}},\ and\ \bibinfo {author} {\bibfnamefont
  {Q.}~\bibnamefont {Lu}},\ }\bibfield  {title} {\enquote {\bibinfo {title}
  {{Atomic imaging with a 12 T magnetic field perpendicular or parallel to the
  sample surface by an ultra-stable scanning tunneling microscope}},}\ }\href
  {https://doi.org/https://doi.org/10.1016/j.ultramic.2023.113774} {\bibfield
  {journal} {\bibinfo  {journal} {Ultramicroscopy}\ }\textbf {\bibinfo {volume}
  {251}},\ \bibinfo {pages} {113774} (\bibinfo {year} {2023})}\BibitemShut
  {NoStop}%
\bibitem [{\citenamefont {Suderow}, \citenamefont {Guillamon},\ and\
  \citenamefont {Vieira}(2011)}]{10.1063/1.3567008}%
  \BibitemOpen
  \bibfield  {author} {\bibinfo {author} {\bibfnamefont {H.}~\bibnamefont
  {Suderow}}, \bibinfo {author} {\bibfnamefont {I.}~\bibnamefont {Guillamon}},\
  and\ \bibinfo {author} {\bibfnamefont {S.}~\bibnamefont {Vieira}},\
  }\bibfield  {title} {\enquote {\bibinfo {title} {Compact very low temperature
  scanning tunneling microscope with mechanically driven horizontal linear
  positioning stage},}\ }\href {https://doi.org/10.1063/1.3567008} {\bibfield
  {journal} {\bibinfo  {journal} {Review of Scientific Instruments}\ }\textbf
  {\bibinfo {volume} {82}},\ \bibinfo {pages} {033711} (\bibinfo {year}
  {2011})}\BibitemShut {NoStop}%
\bibitem [{\citenamefont {Martín-Vega}\ \emph {et~al.}(2021)\citenamefont
  {Martín-Vega}, \citenamefont {Barrena}, \citenamefont {Sánchez-Barquilla},
  \citenamefont {Fernández-Lomana}, \citenamefont {Benito~Llorens},
  \citenamefont {Wu}, \citenamefont {Fente}, \citenamefont {Perconte~Duplain},
  \citenamefont {Horcas}, \citenamefont {López}, \citenamefont {Blanco},
  \citenamefont {Higuera}, \citenamefont {Mañas-Valero}, \citenamefont {Jo},
  \citenamefont {Schmidt}, \citenamefont {Canfield}, \citenamefont
  {Rubio-Bollinger}, \citenamefont {Rodrigo}, \citenamefont {Herrera},
  \citenamefont {Guillamón},\ and\ \citenamefont
  {Suderow}}]{10.1063/5.0064511}%
  \BibitemOpen
  \bibfield  {author} {\bibinfo {author} {\bibfnamefont {F.}~\bibnamefont
  {Martín-Vega}}, \bibinfo {author} {\bibfnamefont {V.}~\bibnamefont
  {Barrena}}, \bibinfo {author} {\bibfnamefont {R.}~\bibnamefont
  {Sánchez-Barquilla}}, \bibinfo {author} {\bibfnamefont {M.}~\bibnamefont
  {Fernández-Lomana}}, \bibinfo {author} {\bibfnamefont {J.}~\bibnamefont
  {Benito~Llorens}}, \bibinfo {author} {\bibfnamefont {B.}~\bibnamefont {Wu}},
  \bibinfo {author} {\bibfnamefont {A.}~\bibnamefont {Fente}}, \bibinfo
  {author} {\bibfnamefont {D.}~\bibnamefont {Perconte~Duplain}}, \bibinfo
  {author} {\bibfnamefont {I.}~\bibnamefont {Horcas}}, \bibinfo {author}
  {\bibfnamefont {R.}~\bibnamefont {López}}, \bibinfo {author} {\bibfnamefont
  {et~al.}},\ }\bibfield  {title} {\enquote {\bibinfo
  {title} {Simplified feedback control system for scanning tunneling
  microscopy},}\ }\href {https://doi.org/10.1063/5.0064511} {\bibfield
  {journal} {\bibinfo  {journal} {Review of Scientific Instruments}\ }\textbf
  {\bibinfo {volume} {92}},\ \bibinfo {pages} {103705} (\bibinfo {year}
  {2021})}\BibitemShut {NoStop}%
\bibitem [{\citenamefont {Pan}, \citenamefont {Hudson},\ and\ \citenamefont
  {Davis}(1999)}]{10.1063/1.1149605}%
  \BibitemOpen
  \bibfield  {author} {\bibinfo {author} {\bibfnamefont {S.~H.}\ \bibnamefont
  {Pan}}, \bibinfo {author} {\bibfnamefont {E.~W.}\ \bibnamefont {Hudson}},\
  and\ \bibinfo {author} {\bibfnamefont {J.~C.}\ \bibnamefont {Davis}},\
  }\bibfield  {title} {\enquote {\bibinfo {title} {$^3${He} refrigerator based
  very low temperature scanning tunneling microscope},}\ }\href
  {https://doi.org/10.1063/1.1149605} {\bibfield  {journal} {\bibinfo
  {journal} {Review of Scientific Instruments}\ }\textbf {\bibinfo {volume}
  {70}},\ \bibinfo {pages} {1459--1463} (\bibinfo {year} {1999})}\BibitemShut
  {NoStop}%
\bibitem [{\citenamefont {Voigtl\"ander}(2015)}]{Voigtlander}%
  \BibitemOpen
  \bibfield  {author} {\bibinfo {author} {\bibfnamefont {B.}~\bibnamefont
  {Voigtl\"ander}},\ }\href {https://doi.org/10.1007/978-3-662-45240-0} {\emph
  {\bibinfo {title} {{Scanning Probe Microscopy: Atomic Force Microscopy and
  Scanning Tunneling Microscopy}}}},\ NanoScience and Technology\ (\bibinfo
  {publisher} {Springer},\ \bibinfo {year} {2015})\BibitemShut {NoStop}%
\bibitem [{\citenamefont {Battisti}\ \emph {et~al.}(2018)\citenamefont
  {Battisti}, \citenamefont {Verdoes}, \citenamefont {van Oosten},
  \citenamefont {Bastiaans},\ and\ \citenamefont {Allan}}]{10.1063/1.5064442}%
  \BibitemOpen
  \bibfield  {author} {\bibinfo {author} {\bibfnamefont {I.}~\bibnamefont
  {Battisti}}, \bibinfo {author} {\bibfnamefont {G.}~\bibnamefont {Verdoes}},
  \bibinfo {author} {\bibfnamefont {K.}~\bibnamefont {van Oosten}}, \bibinfo
  {author} {\bibfnamefont {K.~M.}\ \bibnamefont {Bastiaans}},\ and\ \bibinfo
  {author} {\bibfnamefont {M.~P.}\ \bibnamefont {Allan}},\ }\bibfield  {title}
  {\enquote {\bibinfo {title} {Definition of design guidelines, construction,
  and performance of an ultra-stable scanning tunneling microscope for
  spectroscopic imaging},}\ }\href {https://doi.org/10.1063/1.5064442}
  {\bibfield  {journal} {\bibinfo  {journal} {Review of Scientific
  Instruments}\ }\textbf {\bibinfo {volume} {89}},\ \bibinfo {pages} {123705}
  (\bibinfo {year} {2018})}\BibitemShut {NoStop}%
\bibitem [{\citenamefont {Wong}\ \emph {et~al.}(2020)\citenamefont {Wong},
  \citenamefont {Jeon}, \citenamefont {Nuckolls}, \citenamefont {Oh},
  \citenamefont {Kingsley},\ and\ \citenamefont {Yazdani}}]{10.1063/1.5132872}%
  \BibitemOpen
  \bibfield  {author} {\bibinfo {author} {\bibfnamefont {D.}~\bibnamefont
  {Wong}}, \bibinfo {author} {\bibfnamefont {S.}~\bibnamefont {Jeon}}, \bibinfo
  {author} {\bibfnamefont {K.~P.}\ \bibnamefont {Nuckolls}}, \bibinfo {author}
  {\bibfnamefont {M.}~\bibnamefont {Oh}}, \bibinfo {author} {\bibfnamefont
  {S.~C.~J.}\ \bibnamefont {Kingsley}},\ and\ \bibinfo {author} {\bibfnamefont
  {A.}~\bibnamefont {Yazdani}},\ }\bibfield  {title} {\enquote {\bibinfo
  {title} {A modular ultra-high vacuum millikelvin scanning tunneling
  microscope},}\ }\href {https://doi.org/10.1063/1.5132872} {\bibfield
  {journal} {\bibinfo  {journal} {Review of Scientific Instruments}\ }\textbf
  {\bibinfo {volume} {91}},\ \bibinfo {pages} {023703} (\bibinfo {year}
  {2020})}\BibitemShut {NoStop}%
\bibitem [{\citenamefont {Misra}\ \emph {et~al.}(2013)\citenamefont {Misra},
  \citenamefont {Zhou}, \citenamefont {Drozdov}, \citenamefont {Seo},
  \citenamefont {Urban}, \citenamefont {Gyenis}, \citenamefont {Kingsley},
  \citenamefont {Jones},\ and\ \citenamefont {Yazdani}}]{10.1063/1.4822271}%
  \BibitemOpen
  \bibfield  {author} {\bibinfo {author} {\bibfnamefont {S.}~\bibnamefont
  {Misra}}, \bibinfo {author} {\bibfnamefont {B.~B.}\ \bibnamefont {Zhou}},
  \bibinfo {author} {\bibfnamefont {I.~K.}\ \bibnamefont {Drozdov}}, \bibinfo
  {author} {\bibfnamefont {J.}~\bibnamefont {Seo}}, \bibinfo {author}
  {\bibfnamefont {L.}~\bibnamefont {Urban}}, \bibinfo {author} {\bibfnamefont
  {A.}~\bibnamefont {Gyenis}}, \bibinfo {author} {\bibfnamefont {S.~C.~J.}\
  \bibnamefont {Kingsley}}, \bibinfo {author} {\bibfnamefont {H.}~\bibnamefont
  {Jones}},\ and\ \bibinfo {author} {\bibfnamefont {A.}~\bibnamefont
  {Yazdani}},\ }\bibfield  {title} {\enquote {\bibinfo {title} {Design and
  performance of an ultra-high vacuum scanning tunneling microscope operating
  at dilution refrigerator temperatures and high magnetic fields},}\ }\href
  {https://doi.org/10.1063/1.4822271} {\bibfield  {journal} {\bibinfo
  {journal} {Review of Scientific Instruments}\ }\textbf {\bibinfo {volume}
  {84}},\ \bibinfo {pages} {103903} (\bibinfo {year} {2013})}\BibitemShut
  {NoStop}%
\bibitem [{\citenamefont {Song}\ \emph {et~al.}(2010)\citenamefont {Song},
  \citenamefont {Otte}, \citenamefont {Shvarts}, \citenamefont {Zhao},
  \citenamefont {Kuk}, \citenamefont {Blankenship}, \citenamefont {Band},
  \citenamefont {Hess},\ and\ \citenamefont {Stroscio}}]{10.1063/1.3520482}%
  \BibitemOpen
  \bibfield  {author} {\bibinfo {author} {\bibfnamefont {Y.~J.}\ \bibnamefont
  {Song}}, \bibinfo {author} {\bibfnamefont {A.~F.}\ \bibnamefont {Otte}},
  \bibinfo {author} {\bibfnamefont {V.}~\bibnamefont {Shvarts}}, \bibinfo
  {author} {\bibfnamefont {Z.}~\bibnamefont {Zhao}}, \bibinfo {author}
  {\bibfnamefont {Y.}~\bibnamefont {Kuk}}, \bibinfo {author} {\bibfnamefont
  {S.~R.}\ \bibnamefont {Blankenship}}, \bibinfo {author} {\bibfnamefont
  {A.}~\bibnamefont {Band}}, \bibinfo {author} {\bibfnamefont {F.~M.}\
  \bibnamefont {Hess}},\ and\ \bibinfo {author} {\bibfnamefont {J.~A.}\
  \bibnamefont {Stroscio}},\ }\bibfield  {title} {\enquote {\bibinfo {title}
  {{Invited Review Article: A 10 mK scanning probe microscopy facility}},}\
  }\href {https://doi.org/10.1063/1.3520482} {\bibfield  {journal} {\bibinfo
  {journal} {Review of Scientific Instruments}\ }\textbf {\bibinfo {volume}
  {81}},\ \bibinfo {pages} {121101} (\bibinfo {year} {2010})}\BibitemShut
  {NoStop}%
\bibitem [{\citenamefont {Álvarez Montoya}\ \emph {et~al.}(2019)\citenamefont
  {Álvarez Montoya}, \citenamefont {Delgado}, \citenamefont {Castilla},
  \citenamefont {Navarrete}, \citenamefont {Contreras}, \citenamefont
  {Marijuan}, \citenamefont {Barrena}, \citenamefont {Guillamón},\ and\
  \citenamefont {Suderow}}]{MONTOYA2019e00058}%
  \BibitemOpen
  \bibfield  {author} {\bibinfo {author} {\bibfnamefont {R.}~\bibnamefont
  {Álvarez Montoya}}, \bibinfo {author} {\bibfnamefont {S.}~\bibnamefont
  {Delgado}}, \bibinfo {author} {\bibfnamefont {J.}~\bibnamefont {Castilla}},
  \bibinfo {author} {\bibfnamefont {J.}~\bibnamefont {Navarrete}}, \bibinfo
  {author} {\bibfnamefont {N.~D.}\ \bibnamefont {Contreras}}, \bibinfo {author}
  {\bibfnamefont {J.~R.}\ \bibnamefont {Marijuan}}, \bibinfo {author}
  {\bibfnamefont {V.}~\bibnamefont {Barrena}}, \bibinfo {author} {\bibfnamefont
  {I.}~\bibnamefont {Guillamón}},\ and\ \bibinfo {author} {\bibfnamefont
  {H.}~\bibnamefont {Suderow}},\ }\bibfield  {title} {\enquote {\bibinfo
  {title} {Methods to simplify cooling of liquid {H}elium cryostats},}\ }\href
  {https://doi.org/10.1016/j.ohx.2019.e00058} {\bibfield  {journal} {\bibinfo
  {journal} {HardwareX}\ }\textbf {\bibinfo {volume} {5}},\ \bibinfo {pages}
  {e00058} (\bibinfo {year} {2019})}\BibitemShut {NoStop}%
\bibitem [{\citenamefont {Revolinsky}, \citenamefont {Spiering},\ and\
  \citenamefont {Beerntsen}(1965)}]{REVOLINSKY19651029}%
  \BibitemOpen
  \bibfield  {author} {\bibinfo {author} {\bibfnamefont {E.}~\bibnamefont
  {Revolinsky}}, \bibinfo {author} {\bibfnamefont {G.}~\bibnamefont
  {Spiering}},\ and\ \bibinfo {author} {\bibfnamefont {D.}~\bibnamefont
  {Beerntsen}},\ }\bibfield  {title} {\enquote {\bibinfo {title}
  {Superconductivity in the niobium-selenium system},}\ }\href
  {https://doi.org/https://doi.org/10.1016/0022-3697(65)90190-3} {\bibfield
  {journal} {\bibinfo  {journal} {Journal of Physics and Chemistry of Solids}\
  }\textbf {\bibinfo {volume} {26}},\ \bibinfo {pages} {1029--1034} (\bibinfo
  {year} {1965})}\BibitemShut {NoStop}%
\bibitem [{\citenamefont {Valero}(2021)}]{tesissamuel}%
  \BibitemOpen
  \bibfield  {author} {\bibinfo {author} {\bibfnamefont {S.~M.}\ \bibnamefont
  {Valero}},\ }\href
  {https://www.educacion.gob.es/teseo/imprimirFichaConsulta.do?idFicha=682954}
  {\emph {\bibinfo {title} {Two-dimensional crystals and van der {Waals}
  heterostructures based on inorganic and molecular strongly correlated layered
  materials}}}\ (\bibinfo {year} {2021})\BibitemShut {NoStop}%
\bibitem [{\citenamefont {Rubio-Bollinger}, \citenamefont {Joyez},\ and\
  \citenamefont {Agra\"{\i}t}(2004)}]{PhysRevLett.93.116803}%
  \BibitemOpen
  \bibfield  {author} {\bibinfo {author} {\bibfnamefont {G.}~\bibnamefont
  {Rubio-Bollinger}}, \bibinfo {author} {\bibfnamefont {P.}~\bibnamefont
  {Joyez}},\ and\ \bibinfo {author} {\bibfnamefont {N.}~\bibnamefont
  {Agra\"{\i}t}},\ }\bibfield  {title} {\enquote {\bibinfo {title} {Metallic
  adhesion in atomic-size junctions},}\ }\href
  {https://doi.org/10.1103/PhysRevLett.93.116803} {\bibfield  {journal}
  {\bibinfo  {journal} {Physical Review Letters}\ }\textbf {\bibinfo {volume}
  {93}},\ \bibinfo {pages} {116803} (\bibinfo {year} {2004})}\BibitemShut
  {NoStop}%
\bibitem [{\citenamefont {Agra\"{\i}t}, \citenamefont {Rodrigo},\ and\
  \citenamefont {Vieira}(1993)}]{Agrait1993}%
  \BibitemOpen
  \bibfield  {author} {\bibinfo {author} {\bibfnamefont {N.}~\bibnamefont
  {Agra\"{\i}t}}, \bibinfo {author} {\bibfnamefont {J.~G.}\ \bibnamefont
  {Rodrigo}},\ and\ \bibinfo {author} {\bibfnamefont {S.}~\bibnamefont
  {Vieira}},\ }\bibfield  {title} {\enquote {\bibinfo {title} {Conductance
  steps and quantization in atomic-size contacts},}\ }\href
  {https://doi.org/10.1103/PhysRevB.47.12345} {\bibfield  {journal} {\bibinfo
  {journal} {Physical Review B}\ }\textbf {\bibinfo {volume} {47}},\ \bibinfo
  {pages} {12345--12348} (\bibinfo {year} {1993})}\BibitemShut {NoStop}%
\bibitem [{\citenamefont {Ohnishi}, \citenamefont {Kondo},\ and\ \citenamefont
  {Takayanagi}(1998)}]{Ohnishi1998}%
  \BibitemOpen
  \bibfield  {author} {\bibinfo {author} {\bibfnamefont {H.}~\bibnamefont
  {Ohnishi}}, \bibinfo {author} {\bibfnamefont {Y.}~\bibnamefont {Kondo}},\
  and\ \bibinfo {author} {\bibfnamefont {K.}~\bibnamefont {Takayanagi}},\
  }\bibfield  {title} {\enquote {\bibinfo {title} {{Quantized conductance
  through individual rows of suspended gold atoms}},}\ }\href
  {https://doi.org/10.1038/27399} {\bibfield  {journal} {\bibinfo  {journal}
  {Nature}\ }\textbf {\bibinfo {volume} {395}},\ \bibinfo {pages} {780--783}
  (\bibinfo {year} {1998})}\BibitemShut {NoStop}%
\bibitem [{\citenamefont {Yanson}\ \emph {et~al.}(1998)\citenamefont {Yanson},
  \citenamefont {Bollinger}, \citenamefont {van~den Brom}, \citenamefont
  {Agraït},\ and\ \citenamefont {van Ruitenbeek}}]{Yanson1998}%
  \BibitemOpen
  \bibfield  {author} {\bibinfo {author} {\bibfnamefont {A.~I.}\ \bibnamefont
  {Yanson}}, \bibinfo {author} {\bibfnamefont {G.~R.}\ \bibnamefont
  {Bollinger}}, \bibinfo {author} {\bibfnamefont {H.~E.}\ \bibnamefont {van~den
  Brom}}, \bibinfo {author} {\bibfnamefont {N.}~\bibnamefont {Agraït}},\ and\
  \bibinfo {author} {\bibfnamefont {J.~M.}\ \bibnamefont {van Ruitenbeek}},\
  }\bibfield  {title} {\enquote {\bibinfo {title} {{Formation and manipulation
  of a metallic wire of single gold atoms}},}\ }\href
  {https://doi.org/10.1038/27405} {\bibfield  {journal} {\bibinfo  {journal}
  {Nature}\ }\textbf {\bibinfo {volume} {395}},\ \bibinfo {pages} {783--785}
  (\bibinfo {year} {1998})}\BibitemShut {NoStop}%
\bibitem [{\citenamefont {Scheer}\ \emph {et~al.}(1998)\citenamefont {Scheer},
  \citenamefont {Agra\"{i}t}, \citenamefont {Cuevas}, \citenamefont {Yeyati},
  \citenamefont {Ludoph}, \citenamefont {Martín-Rodero}, \citenamefont
  {Bollinger}, \citenamefont {van Ruitenbeek},\ and\ \citenamefont
  {Urbina}}]{Scheer1998}%
  \BibitemOpen
  \bibfield  {author} {\bibinfo {author} {\bibfnamefont {E.}~\bibnamefont
  {Scheer}}, \bibinfo {author} {\bibfnamefont {N.}~\bibnamefont {Agra\"{i}t}},
  \bibinfo {author} {\bibfnamefont {J.~C.}\ \bibnamefont {Cuevas}}, \bibinfo
  {author} {\bibfnamefont {A.~L.}\ \bibnamefont {Yeyati}}, \bibinfo {author}
  {\bibfnamefont {B.}~\bibnamefont {Ludoph}}, \bibinfo {author} {\bibfnamefont
  {A.}~\bibnamefont {Martín-Rodero}}, \bibinfo {author} {\bibfnamefont
  {G.~R.}\ \bibnamefont {Bollinger}}, \bibinfo {author} {\bibfnamefont {J.~M.}\
  \bibnamefont {van Ruitenbeek}},\ and\ \bibinfo {author} {\bibfnamefont
  {C.}~\bibnamefont {Urbina}},\ }\bibfield  {title} {\enquote {\bibinfo {title}
  {{The signature of chemical valence in the electrical conduction through a
  single-atom contact}},}\ }\href {https://doi.org/10.1038/28112} {\bibfield
  {journal} {\bibinfo  {journal} {Nature}\ }\textbf {\bibinfo {volume} {394}},\
  \bibinfo {pages} {154--157} (\bibinfo {year} {1998})}\BibitemShut {NoStop}%
\bibitem [{\citenamefont {Rodrigues}, \citenamefont {Fuhrer},\ and\
  \citenamefont {Ugarte}(2000)}]{Rodrigues2000Au}%
  \BibitemOpen
  \bibfield  {author} {\bibinfo {author} {\bibfnamefont {V.}~\bibnamefont
  {Rodrigues}}, \bibinfo {author} {\bibfnamefont {T.}~\bibnamefont {Fuhrer}},\
  and\ \bibinfo {author} {\bibfnamefont {D.}~\bibnamefont {Ugarte}},\
  }\bibfield  {title} {\enquote {\bibinfo {title} {Signature of atomic
  structure in the quantum conductance of gold nanowires},}\ }\href
  {https://doi.org/10.1103/PhysRevLett.85.4124} {\bibfield  {journal} {\bibinfo
   {journal} {Physical Review Letters}\ }\textbf {\bibinfo {volume} {85}},\
  \bibinfo {pages} {4124--4127} (\bibinfo {year} {2000})}\BibitemShut {NoStop}%
\bibitem [{\citenamefont {Krans}\ \emph {et~al.}(1995)\citenamefont {Krans},
  \citenamefont {van Ruitenbeek}, \citenamefont {Fisun}, \citenamefont
  {Yanson},\ and\ \citenamefont {de~Jongh}}]{Krans1995}%
  \BibitemOpen
  \bibfield  {author} {\bibinfo {author} {\bibfnamefont {J.~M.}\ \bibnamefont
  {Krans}}, \bibinfo {author} {\bibfnamefont {J.~M.}\ \bibnamefont {van
  Ruitenbeek}}, \bibinfo {author} {\bibfnamefont {V.~V.}\ \bibnamefont
  {Fisun}}, \bibinfo {author} {\bibfnamefont {I.~K.}\ \bibnamefont {Yanson}},\
  and\ \bibinfo {author} {\bibfnamefont {L.~J.}\ \bibnamefont {de~Jongh}},\
  }\bibfield  {title} {\enquote {\bibinfo {title} {The signature of conductance
  quantization in metallic point contacts},}\ }\href
  {https://doi.org/10.1038/375767a0} {\bibfield  {journal} {\bibinfo  {journal}
  {Nature}\ }\textbf {\bibinfo {volume} {375}},\ \bibinfo {pages} {767--769}
  (\bibinfo {year} {1995})}\BibitemShut {NoStop}%
\bibitem [{\citenamefont {Agra\"{i}t}, \citenamefont {Yeyati},\ and\
  \citenamefont {{van Ruitenbeek}}(2003)}]{AGRAIT2003}%
  \BibitemOpen
  \bibfield  {author} {\bibinfo {author} {\bibfnamefont {N.}~\bibnamefont
  {Agra\"{i}t}}, \bibinfo {author} {\bibfnamefont {A.~L.}\ \bibnamefont
  {Yeyati}},\ and\ \bibinfo {author} {\bibfnamefont {J.~M.}\ \bibnamefont {{van
  Ruitenbeek}}},\ }\bibfield  {title} {\enquote {\bibinfo {title} {{Quantum
  properties of atomic-sized conductors}},}\ }\href
  {https://doi.org/https://doi.org/10.1016/S0370-1573(02)00633-6} {\bibfield
  {journal} {\bibinfo  {journal} {Physics Reports}\ }\textbf {\bibinfo {volume}
  {377}},\ \bibinfo {pages} {81--279} (\bibinfo {year} {2003})}\BibitemShut
  {NoStop}%
\bibitem [{\citenamefont {Sirvent}\ \emph {et~al.}(1996)\citenamefont
  {Sirvent}, \citenamefont {Rodrigo}, \citenamefont {Vieira}, \citenamefont
  {Jurczyszyn}, \citenamefont {Mingo},\ and\ \citenamefont
  {Flores}}]{Sirvent1996}%
  \BibitemOpen
  \bibfield  {author} {\bibinfo {author} {\bibfnamefont {C.}~\bibnamefont
  {Sirvent}}, \bibinfo {author} {\bibfnamefont {J.~G.}\ \bibnamefont
  {Rodrigo}}, \bibinfo {author} {\bibfnamefont {S.}~\bibnamefont {Vieira}},
  \bibinfo {author} {\bibfnamefont {L.}~\bibnamefont {Jurczyszyn}}, \bibinfo
  {author} {\bibfnamefont {N.}~\bibnamefont {Mingo}},\ and\ \bibinfo {author}
  {\bibfnamefont {F.}~\bibnamefont {Flores}},\ }\bibfield  {title} {\enquote
  {\bibinfo {title} {{Conductance step for a single-atom contact in the
  scanning tunneling microscope: Noble and transition metals}},}\ }\href
  {https://doi.org/10.1103/PhysRevB.53.16086} {\bibfield  {journal} {\bibinfo
  {journal} {Physical Review B}\ }\textbf {\bibinfo {volume} {53}},\ \bibinfo
  {pages} {16086--16090} (\bibinfo {year} {1996})}\BibitemShut {NoStop}%
\bibitem [{\citenamefont {Sabater}\ \emph {et~al.}(2012)\citenamefont
  {Sabater}, \citenamefont {Untiedt}, \citenamefont {Palacios},\ and\
  \citenamefont {Caturla}}]{PhysRevLett.108.205502}%
  \BibitemOpen
  \bibfield  {author} {\bibinfo {author} {\bibfnamefont {C.}~\bibnamefont
  {Sabater}}, \bibinfo {author} {\bibfnamefont {C.}~\bibnamefont {Untiedt}},
  \bibinfo {author} {\bibfnamefont {J.~J.}\ \bibnamefont {Palacios}},\ and\
  \bibinfo {author} {\bibfnamefont {M.~J.}\ \bibnamefont {Caturla}},\
  }\bibfield  {title} {\enquote {\bibinfo {title} {{Mechanical annealing of
  metallic electrodes at the atomic scale}},}\ }\href
  {https://doi.org/10.1103/PhysRevLett.108.205502} {\bibfield  {journal}
  {\bibinfo  {journal} {Physical Review Letters}\ }\textbf {\bibinfo {volume}
  {108}},\ \bibinfo {pages} {205502} (\bibinfo {year} {2012})}\BibitemShut
  {NoStop}%
\bibitem [{\citenamefont {Sabater}\ \emph {et~al.}(2018)\citenamefont
  {Sabater}, \citenamefont {Dednam}, \citenamefont {Calvo}, \citenamefont
  {Fern\'andez}, \citenamefont {Untiedt},\ and\ \citenamefont
  {Caturla}}]{Sabater2018}%
  \BibitemOpen
  \bibfield  {author} {\bibinfo {author} {\bibfnamefont {C.}~\bibnamefont
  {Sabater}}, \bibinfo {author} {\bibfnamefont {W.}~\bibnamefont {Dednam}},
  \bibinfo {author} {\bibfnamefont {M.~R.}\ \bibnamefont {Calvo}}, \bibinfo
  {author} {\bibfnamefont {M.~A.}\ \bibnamefont {Fern\'andez}}, \bibinfo
  {author} {\bibfnamefont {C.}~\bibnamefont {Untiedt}},\ and\ \bibinfo {author}
  {\bibfnamefont {M.~J.}\ \bibnamefont {Caturla}},\ }\bibfield  {title}
  {\enquote {\bibinfo {title} {{Role of first-neighbor geometry in the
  electronic and mechanical properties of atomic contacts}},}\ }\href
  {https://doi.org/10.1103/PhysRevB.97.075418} {\bibfield  {journal} {\bibinfo
  {journal} {Physical Review B}\ }\textbf {\bibinfo {volume} {97}},\ \bibinfo
  {pages} {075418} (\bibinfo {year} {2018})}\BibitemShut {NoStop}%
\bibitem [{\citenamefont {Rodrigo}\ \emph {et~al.}(2004)\citenamefont
  {Rodrigo}, \citenamefont {Suderow}, \citenamefont {Vieira}, \citenamefont
  {Bascones},\ and\ \citenamefont {Guinea}}]{JGRodrigo_2004}%
  \BibitemOpen
  \bibfield  {author} {\bibinfo {author} {\bibfnamefont {J.~G.}\ \bibnamefont
  {Rodrigo}}, \bibinfo {author} {\bibfnamefont {H.}~\bibnamefont {Suderow}},
  \bibinfo {author} {\bibfnamefont {S.}~\bibnamefont {Vieira}}, \bibinfo
  {author} {\bibfnamefont {E.}~\bibnamefont {Bascones}},\ and\ \bibinfo
  {author} {\bibfnamefont {F.}~\bibnamefont {Guinea}},\ }\bibfield  {title}
  {\enquote {\bibinfo {title} {Superconducting nanostructures fabricated with
  the scanning tunnelling microscope},}\ }\href
  {https://doi.org/10.1088/0953-8984/16/34/R01} {\bibfield  {journal} {\bibinfo
   {journal} {Journal of Physics: Condensed Matter}\ }\textbf {\bibinfo
  {volume} {16}},\ \bibinfo {pages} {R1151} (\bibinfo {year}
  {2004})}\BibitemShut {NoStop}%
\bibitem [{\citenamefont {Cuevas}\ and\ \citenamefont
  {Scheer}(2017)}]{bookCuevasScheer}%
  \BibitemOpen
  \bibfield  {author} {\bibinfo {author} {\bibfnamefont {J.~C.}\ \bibnamefont
  {Cuevas}}\ and\ \bibinfo {author} {\bibfnamefont {E.}~\bibnamefont
  {Scheer}},\ }\href
  {https://www.worldscientific.com/worldscibooks/10.1142/10598} {\emph
  {\bibinfo {title} {Molecular Electronics, an introduction to theory and
  experiment}}},\ \bibinfo {edition} {2nd}\ ed.\ (\bibinfo  {publisher} {World
  Scientific},\ \bibinfo {year} {2017})\BibitemShut {NoStop}%
\bibitem [{\citenamefont {Requist}\ \emph {et~al.}(2016)\citenamefont
  {Requist}, \citenamefont {Baruselli}, \citenamefont {Smogunov}, \citenamefont
  {Fabrizio}, \citenamefont {Modesti},\ and\ \citenamefont
  {Tosatti}}]{Requist2016}%
  \BibitemOpen
  \bibfield  {author} {\bibinfo {author} {\bibfnamefont {R.}~\bibnamefont
  {Requist}}, \bibinfo {author} {\bibfnamefont {P.~P.}\ \bibnamefont
  {Baruselli}}, \bibinfo {author} {\bibfnamefont {A.}~\bibnamefont {Smogunov}},
  \bibinfo {author} {\bibfnamefont {M.}~\bibnamefont {Fabrizio}}, \bibinfo
  {author} {\bibfnamefont {S.}~\bibnamefont {Modesti}},\ and\ \bibinfo {author}
  {\bibfnamefont {E.}~\bibnamefont {Tosatti}},\ }\bibfield  {title} {\enquote
  {\bibinfo {title} {{Metallic, magnetic and molecular nanocontacts}},}\ }\href
  {https://doi.org/10.1038/nnano.2016.55} {\bibfield  {journal} {\bibinfo
  {journal} {Nature Nanotechnology}\ }\textbf {\bibinfo {volume} {11}},\
  \bibinfo {pages} {499--508} (\bibinfo {year} {2016})}\BibitemShut {NoStop}%
\bibitem [{\citenamefont {Campbell}, \citenamefont {Doria},\ and\ \citenamefont
  {Kogan}(1988)}]{PRB.38.4}%
  \BibitemOpen
  \bibfield  {author} {\bibinfo {author} {\bibfnamefont {L.~J.}\ \bibnamefont
  {Campbell}}, \bibinfo {author} {\bibfnamefont {M.~M.}\ \bibnamefont
  {Doria}},\ and\ \bibinfo {author} {\bibfnamefont {V.~G.}\ \bibnamefont
  {Kogan}},\ }\bibfield  {title} {\enquote {\bibinfo {title} {{Vortex lattice
  structures in uniaxial superconductors}},}\ }\href
  {https://doi.org/10.1103/PhysRevB.38.2439} {\bibfield  {journal} {\bibinfo
  {journal} {Physical Review B}\ }\textbf {\bibinfo {volume} {38}},\ \bibinfo
  {pages} {2339} (\bibinfo {year} {1988})}\BibitemShut {NoStop}%
\bibitem [{\citenamefont {Dvir}\ \emph {et~al.}(2018)\citenamefont {Dvir},
  \citenamefont {Massee}, \citenamefont {Attias}, \citenamefont {Khodas},
  \citenamefont {Aprili}, \citenamefont {Quay},\ and\ \citenamefont
  {Steinberg}}]{Dvir2018}%
  \BibitemOpen
  \bibfield  {author} {\bibinfo {author} {\bibfnamefont {T.}~\bibnamefont
  {Dvir}}, \bibinfo {author} {\bibfnamefont {F.}~\bibnamefont {Massee}},
  \bibinfo {author} {\bibfnamefont {L.}~\bibnamefont {Attias}}, \bibinfo
  {author} {\bibfnamefont {M.}~\bibnamefont {Khodas}}, \bibinfo {author}
  {\bibfnamefont {M.}~\bibnamefont {Aprili}}, \bibinfo {author} {\bibfnamefont
  {C.~H.~L.}\ \bibnamefont {Quay}},\ and\ \bibinfo {author} {\bibfnamefont
  {H.}~\bibnamefont {Steinberg}},\ }\bibfield  {title} {\enquote {\bibinfo
  {title} {{Spectroscopy of bulk and few-layer superconducting
  ${\mathrm{NbSe}}_{2}$ with van der Waals tunnel junctions}},}\ }\href
  {https://doi.org/10.1038/s41467-018-03000-w} {\bibfield  {journal} {\bibinfo
  {journal} {Nature Communications}\ }\textbf {\bibinfo {volume} {9}},\
  \bibinfo {pages} {598} (\bibinfo {year} {2018})}\BibitemShut {NoStop}%
\bibitem [{\citenamefont {Kuzmanovi\ifmmode~\acute{c}\else \'{c}\fi{}}\ \emph
  {et~al.}(2022)\citenamefont {Kuzmanovi\ifmmode~\acute{c}\else \'{c}\fi{}},
  \citenamefont {Dvir}, \citenamefont {LeBoeuf}, \citenamefont
  {Ili\ifmmode~\acute{c}\else \'{c}\fi{}}, \citenamefont {Haim}, \citenamefont
  {M\"ockli}, \citenamefont {Kramer}, \citenamefont {Khodas}, \citenamefont
  {Houzet}, \citenamefont {Meyer}, \citenamefont {Aprili}, \citenamefont
  {Steinberg},\ and\ \citenamefont {Quay}}]{PhysRevB.106.184514}%
  \BibitemOpen
  \bibfield  {author} {\bibinfo {author} {\bibfnamefont {M.}~\bibnamefont
  {Kuzmanovi\ifmmode~\acute{c}\else \'{c}\fi{}}}, \bibinfo {author}
  {\bibfnamefont {T.}~\bibnamefont {Dvir}}, \bibinfo {author} {\bibfnamefont
  {D.}~\bibnamefont {LeBoeuf}}, \bibinfo {author} {\bibfnamefont
  {S.}~\bibnamefont {Ili\ifmmode~\acute{c}\else \'{c}\fi{}}}, \bibinfo {author}
  {\bibfnamefont {M.}~\bibnamefont {Haim}}, \bibinfo {author} {\bibfnamefont
  {D.}~\bibnamefont {M\"ockli}}, \bibinfo {author} {\bibfnamefont
  {S.}~\bibnamefont {Kramer}}, \bibinfo {author} {\bibfnamefont
  {M.}~\bibnamefont {Khodas}}, \bibinfo {author} {\bibfnamefont
  {M.}~\bibnamefont {Houzet}}, \bibinfo {author} {\bibfnamefont {J.~S.}\
  \bibnamefont {Meyer}}, \bibinfo {author} {\bibfnamefont {et~al.}},\
  }\bibfield  {title} {\enquote {\bibinfo {title} {{Tunneling spectroscopy of
  few-monolayer ${\mathrm{NbSe}}_{2}$ in high magnetic fields: Triplet
  superconductivity and Ising protection}},}\ }\href
  {https://doi.org/10.1103/PhysRevB.106.184514} {\bibfield  {journal} {\bibinfo
   {journal} {Phys. Rev. B}\ }\textbf {\bibinfo {volume} {106}},\ \bibinfo
  {pages} {184514} (\bibinfo {year} {2022})}\BibitemShut {NoStop}%
\bibitem [{\citenamefont {Foner}\ and\ \citenamefont
  {McNiff}(1973)}]{FONER1973429}%
  \BibitemOpen
  \bibfield  {author} {\bibinfo {author} {\bibfnamefont {S.}~\bibnamefont
  {Foner}}\ and\ \bibinfo {author} {\bibfnamefont {E.}~\bibnamefont {McNiff}},\
  }\bibfield  {title} {\enquote {\bibinfo {title} {{Upper critical fields of
  layered superconducting ${\mathrm{NbSe}}_{2}$ at low temperature}},}\ }\href
  {https://doi.org/https://doi.org/10.1016/0375-9601(73)90693-2} {\bibfield
  {journal} {\bibinfo  {journal} {Physics Letters A}\ }\textbf {\bibinfo
  {volume} {45}},\ \bibinfo {pages} {429--430} (\bibinfo {year}
  {1973})}\BibitemShut {NoStop}%
\bibitem [{\citenamefont {Cho}\ \emph {et~al.}(2022)\citenamefont {Cho},
  \citenamefont {Lyu}, \citenamefont {An}, \citenamefont {Han}, \citenamefont
  {Lo}, \citenamefont {Ng}, \citenamefont {Hu}, \citenamefont {Gao},
  \citenamefont {Li}, \citenamefont {et al.}, \citenamefont {Wang},
  \citenamefont {Schmalian},\ and\ \citenamefont
  {Lortz}}]{PhysRevLett.129.087002}%
  \BibitemOpen
  \bibfield  {author} {\bibinfo {author} {\bibfnamefont {C.-w.}\ \bibnamefont
  {Cho}}, \bibinfo {author} {\bibfnamefont {J.}~\bibnamefont {Lyu}}, \bibinfo
  {author} {\bibfnamefont {L.}~\bibnamefont {An}}, \bibinfo {author}
  {\bibfnamefont {T.}~\bibnamefont {Han}}, \bibinfo {author} {\bibfnamefont
  {K.~T.}\ \bibnamefont {Lo}}, \bibinfo {author} {\bibfnamefont {C.~Y.}\
  \bibnamefont {Ng}}, \bibinfo {author} {\bibfnamefont {J.}~\bibnamefont {Hu}},
  \bibinfo {author} {\bibfnamefont {Y.}~\bibnamefont {Gao}}, \bibinfo {author}
  {\bibfnamefont {G.}~\bibnamefont {Li}}, \bibinfo {author} {\bibfnamefont
  {M.}~\bibnamefont {Huang}}, \bibinfo {author} {\bibfnamefont
  {et~al.}},\ }\bibfield  {title} {\enquote {\bibinfo {title} {{Nodal and
  nematic superconducting phases in ${\mathrm{NbSe}}_{2}$ monolayers from
  competing superconducting channels}},}\ }\href
  {https://doi.org/10.1103/PhysRevLett.129.087002} {\bibfield  {journal}
  {\bibinfo  {journal} {Phys. Rev. Lett.}\ }\textbf {\bibinfo {volume} {129}},\
  \bibinfo {pages} {087002} (\bibinfo {year} {2022})}\BibitemShut {NoStop}%
\bibitem [{\citenamefont {Xi}\ \emph {et~al.}(2016)\citenamefont {Xi},
  \citenamefont {Wang}, \citenamefont {Zhao}, \citenamefont {Park},
  \citenamefont {Law}, \citenamefont {Berger}, \citenamefont {Forr{\'o}},
  \citenamefont {Shan},\ and\ \citenamefont {Mak}}]{Xi2016}%
  \BibitemOpen
  \bibfield  {author} {\bibinfo {author} {\bibfnamefont {X.}~\bibnamefont
  {Xi}}, \bibinfo {author} {\bibfnamefont {Z.}~\bibnamefont {Wang}}, \bibinfo
  {author} {\bibfnamefont {W.}~\bibnamefont {Zhao}}, \bibinfo {author}
  {\bibfnamefont {J.-H.}\ \bibnamefont {Park}}, \bibinfo {author}
  {\bibfnamefont {K.~T.}\ \bibnamefont {Law}}, \bibinfo {author} {\bibfnamefont
  {H.}~\bibnamefont {Berger}}, \bibinfo {author} {\bibfnamefont
  {L.}~\bibnamefont {Forr{\'o}}}, \bibinfo {author} {\bibfnamefont
  {J.}~\bibnamefont {Shan}},\ and\ \bibinfo {author} {\bibfnamefont {K.~F.}\
  \bibnamefont {Mak}},\ }\bibfield  {title} {\enquote {\bibinfo {title} {{Ising
  pairing in superconducting ${\mathrm{NbSe}}_{2}$ atomic layers}},}\ }\href
  {https://doi.org/10.1038/nphys3538} {\bibfield  {journal} {\bibinfo
  {journal} {Nature Physics}\ }\textbf {\bibinfo {volume} {12}},\ \bibinfo
  {pages} {139--143} (\bibinfo {year} {2016})}\BibitemShut {NoStop}%
\bibitem [{\citenamefont {Nader}\ and\ \citenamefont
  {Monceau}(2014)}]{Nader2014}%
  \BibitemOpen
  \bibfield  {author} {\bibinfo {author} {\bibfnamefont {A.}~\bibnamefont
  {Nader}}\ and\ \bibinfo {author} {\bibfnamefont {P.}~\bibnamefont
  {Monceau}},\ }\bibfield  {title} {\enquote {\bibinfo {title} {{Critical field
  of 2H-${\mathrm{NbSe}}_{2}$ down to 50 mK}},}\ }\href
  {https://doi.org/10.1186/2193-1801-3-16} {\bibfield  {journal} {\bibinfo
  {journal} {SpringerPlus}\ }\textbf {\bibinfo {volume} {3}},\ \bibinfo {pages}
  {16} (\bibinfo {year} {2014})}\BibitemShut {NoStop}%
\bibitem [{\citenamefont {Li}\ \emph {et~al.}(2016)\citenamefont {Li} \emph
  {et~al.}}]{Li2016}%
  \BibitemOpen
  \bibfield  {author} {\bibinfo {author} {\bibfnamefont {Y.}~\bibnamefont
  {Li}}\ and\ \bibinfo {author} {\bibfnamefont {G.}~\bibnamefont
  {Kang}}\ and\ \bibinfo {author} {\bibfnamefont {Y.}~\bibnamefont
  {Gao}}}\bibfield  {title} {\enquote {\bibinfo {title} {{Scaling
  rules for critical current density in anisotropic biaxial
  superconductors}},}\ }\href {https://doi.org/10.1016/j.physb.2016.03.027}
  {\bibfield  {journal} {\bibinfo  {journal}
  {\href{http://dx.doi.org/10.1016/j.physb.2016.03.027}{Physica B: Condensed
  Matter}}\ }\textbf {\bibinfo {volume} {491}},\ \bibinfo {pages} {70–78}
  (\bibinfo {year} {2016})}\BibitemShut {NoStop}%
\bibitem [{\citenamefont {Mikitik}\ and\ \citenamefont
  {Brandt}(2009)}]{Mikitik2009}%
  \BibitemOpen
  \bibfield  {author} {\bibinfo {author} {\bibfnamefont {G.~P.}\ \bibnamefont
  {Mikitik}}\ and\ \bibinfo {author} {\bibfnamefont {E.~H.}\ \bibnamefont
  {Brandt}},\ }\bibfield  {title} {\enquote {\bibinfo {title} {{Flux-line
  pinning by point defects in anisotropic biaxial type-II superconductors}},}\
  }\href {https://doi.org/10.1103/PhysRevB.79.020506} {\bibfield  {journal}
  {\bibinfo  {journal} {Physical Review B}\ }\textbf {\bibinfo {volume} {79}},\
  \bibinfo {pages} {020506} (\bibinfo {year} {2009})}\BibitemShut {NoStop}%
\bibitem [{\citenamefont {Berm{\'u}dez-P{\'e}rez}\ \emph
  {et~al.}(2025)\citenamefont {Berm{\'u}dez-P{\'e}rez} \emph
  {et~al.}}]{BermudezPerez2025_FeSeTiltedFields}%
  \BibitemOpen
  \bibfield  {author} {\bibinfo {author} {\bibfnamefont {J.D.}\ \bibnamefont
  {Bermudez-Perez}}, \bibinfo {author} {\bibfnamefont {P.}~\bibnamefont {Garc\'ia Talavera}}, \bibinfo
  {author} {\bibfnamefont {J.A.}~\bibnamefont {Moreno}}, \bibinfo {author}
  {\bibfnamefont {M.}~\bibnamefont {\'Agueda}}, \bibinfo {author} {\bibfnamefont
  {O.}\ \bibnamefont {Bou Marqu\'es}}, \bibinfo {author} {\bibfnamefont {Juan}\
  \bibnamefont {Schmidt}}, \bibinfo {author} {\bibfnamefont {Sergey L.}~\bibnamefont {Bud'ko}},
  \bibinfo {author} {\bibfnamefont {P.C.}~\bibnamefont {Canfield}}, \bibinfo {author}
  {\bibfnamefont {I.}~\bibnamefont {Guillam\'on}}, \bibinfo {author} {\bibfnamefont
  {J.A.}~\bibnamefont {Galvis}}, \bibinfo {author} {\bibfnamefont
  {et~al.}},\ }\href@noop {} {\enquote {\bibinfo
  {title} {{Biaxial Superconductivity in FeSe}},}\ } (\bibinfo {year} {2025}),\
  \bibinfo {note} {manuscript in preparation}\BibitemShut {NoStop}%
\bibitem [{\citenamefont {Sanchez}\ \emph {et~al.}(1995)\citenamefont
  {Sanchez}, \citenamefont {Junod}, \citenamefont {Muller}, \citenamefont
  {Berger},\ and\ \citenamefont {Lévy}}]{PhysB.204.1}%
  \BibitemOpen
  \bibfield  {author} {\bibinfo {author} {\bibfnamefont {D.}~\bibnamefont
  {Sanchez}}, \bibinfo {author} {\bibfnamefont {A.}~\bibnamefont {Junod}},
  \bibinfo {author} {\bibfnamefont {J.}~\bibnamefont {Muller}}, \bibinfo
  {author} {\bibfnamefont {H.}~\bibnamefont {Berger}},\ and\ \bibinfo {author}
  {\bibfnamefont {F.}~\bibnamefont {Lévy}},\ }\bibfield  {title} {\enquote
  {\bibinfo {title} {Specific heat of 2{H}-{${\mathrm{NbSe}}_{2}$} in high
  magnetic fields},}\ }\href {https://doi.org/10.1016/0921-4526(94)00259-X}
  {\bibfield  {journal} {\bibinfo  {journal} {Physica B: Condensed Matter}\
  }\textbf {\bibinfo {volume} {204}},\ \bibinfo {pages} {167--175} (\bibinfo
  {year} {1995})}\BibitemShut {NoStop}%
\bibitem [{\citenamefont {Meng}\ \emph {et~al.}(2020)\citenamefont {Meng},
  \citenamefont {Zhao}, \citenamefont {Wang}, \citenamefont {Zhang},
  \citenamefont {Feng}, \citenamefont {Wang}, \citenamefont {Geng},
  \citenamefont {Guo}, \citenamefont {Hou}, \citenamefont {Pi}, \citenamefont
  {Lu},\ and\ \citenamefont {Lu}}]{MENG2020112975}%
  \BibitemOpen
  \bibfield  {author} {\bibinfo {author} {\bibfnamefont {W.}~\bibnamefont
  {Meng}}, \bibinfo {author} {\bibfnamefont {K.}~\bibnamefont {Zhao}}, \bibinfo
  {author} {\bibfnamefont {J.}~\bibnamefont {Wang}}, \bibinfo {author}
  {\bibfnamefont {J.}~\bibnamefont {Zhang}}, \bibinfo {author} {\bibfnamefont
  {Q.}~\bibnamefont {Feng}}, \bibinfo {author} {\bibfnamefont {Z.}~\bibnamefont
  {Wang}}, \bibinfo {author} {\bibfnamefont {T.}~\bibnamefont {Geng}}, \bibinfo
  {author} {\bibfnamefont {T.}~\bibnamefont {Guo}}, \bibinfo {author}
  {\bibfnamefont {Y.}~\bibnamefont {Hou}}, \bibinfo {author} {\bibfnamefont
  {L.}~\bibnamefont {Pi}}, \bibinfo {author} {\bibfnamefont {Y.}~\bibnamefont
  {Lu}},\ and\ \bibinfo {author} {\bibfnamefont {Q.}~\bibnamefont {Lu}},\
  }\bibfield  {title} {\enquote {\bibinfo {title} {{30 T scanning tunneling
  microscope in a hybrid magnet with essentially non-metallic design}},}\
  }\href {https://doi.org/10.1016/j.ultramic.2020.112975} {\bibfield  {journal}
  {\bibinfo  {journal} {Ultramicroscopy}\ }\textbf {\bibinfo {volume} {212}},\
  \bibinfo {pages} {112975} (\bibinfo {year} {2020})}\BibitemShut {NoStop}%
\bibitem [{\citenamefont {Tao}\ \emph {et~al.}(2017)\citenamefont {Tao},
  \citenamefont {Singh}, \citenamefont {Rossi}, \citenamefont {Gerritsen},
  \citenamefont {Hendriksen}, \citenamefont {Khajetoorians}, \citenamefont
  {Christianen}, \citenamefont {Maan}, \citenamefont {Zeitler},\ and\
  \citenamefont {Bryant}}]{10.1063/1.4995372}%
  \BibitemOpen
  \bibfield  {author} {\bibinfo {author} {\bibfnamefont {W.}~\bibnamefont
  {Tao}}, \bibinfo {author} {\bibfnamefont {S.}~\bibnamefont {Singh}}, \bibinfo
  {author} {\bibfnamefont {L.}~\bibnamefont {Rossi}}, \bibinfo {author}
  {\bibfnamefont {J.~W.}\ \bibnamefont {Gerritsen}}, \bibinfo {author}
  {\bibfnamefont {B.~L.~M.}\ \bibnamefont {Hendriksen}}, \bibinfo {author}
  {\bibfnamefont {A.~A.}\ \bibnamefont {Khajetoorians}}, \bibinfo {author}
  {\bibfnamefont {P.~C.~M.}\ \bibnamefont {Christianen}}, \bibinfo {author}
  {\bibfnamefont {J.~C.}\ \bibnamefont {Maan}}, \bibinfo {author}
  {\bibfnamefont {U.}~\bibnamefont {Zeitler}},\ and\ \bibinfo {author}
  {\bibfnamefont {B.}~\bibnamefont {Bryant}},\ }\bibfield  {title} {\enquote
  {\bibinfo {title} {{A low-temperature scanning tunneling microscope capable
  of microscopy and spectroscopy in a Bitter magnet at up to 34 T}},}\ }\href
  {https://doi.org/10.1063/1.4995372} {\bibfield  {journal} {\bibinfo
  {journal} {Review of Scientific Instruments}\ }\textbf {\bibinfo {volume}
  {88}},\ \bibinfo {pages} {093706} (\bibinfo {year} {2017})}\BibitemShut
  {NoStop}%
\bibitem {NdBi}%
  \BibitemOpen
  \bibfield  {author} {\bibinfo {author} {\bibfnamefont {Benjamin}~\bibnamefont
  {Schrunk}}, \bibinfo {author} {\bibfnamefont {et.}~\bibnamefont {al.}}\ }\bibfield  {title} {\enquote
  {\bibinfo {title} {{Emergence of Fermi arcs due to magnetic splitting in an antiferromagnet}},}\ }\href
  {https://doi.org/10.1038/s41586-022-04412-x} {\bibfield  {journal} {\bibinfo
  {journal} {Nature}\ }\textbf {\bibinfo {volume}
  {603}},\ \bibinfo {pages} {610} (\bibinfo {year} {2022})}\BibitemShut
  {NoStop}%
  \bibitem {Pfleiderer}%
  \BibitemOpen
  \bibfield  {author} {\bibinfo {author} {\bibfnamefont {Christian}~\bibnamefont
  {Pfleiderer}}\ }\bibfield  {title} {\enquote
  {\bibinfo {title} {{Superconducting phases of f-electron compounds}},}\ }\href
  {https://doi.org/10.1103/RevModPhys.81.1551} {\bibfield  {journal} {\bibinfo
  {journal} {Reviews of Modern Physics}\ }\textbf {\bibinfo {volume}
  {81}},\ \bibinfo {pages} {1551} (\bibinfo {year} {2009})}\BibitemShut
  {NoStop}%
  \bibitem {Lewin}%
  \BibitemOpen
  \bibfield  {author} {\bibinfo {author} {\bibfnamefont {Sylvia~K}~\bibnamefont
  {Lewin}}, \bibinfo {author} {\bibfnamefont {Corey~E}~\bibnamefont
  {Frank}}, \bibinfo {author} {\bibfnamefont {Sheng}~\bibnamefont
  {Ran}}, \bibinfo {author} {\bibfnamefont {Johnpierre}~\bibnamefont
  {Paglione}}, and \bibinfo {author} {\bibfnamefont {Nicholas P}~\bibnamefont
  {Butch}}\ }\bibfield  {title} {\enquote
  {\bibinfo {title} {{A review of UTe$_2$ at high magnetic fields}},}\ }\href
  {https://doi.org/10.1088/1361-6633/acfb93} {\bibfield  {journal} {\bibinfo
  {journal} {Reports on Progress in Physics}\ }\textbf {\bibinfo {volume}
  {81}},\ \bibinfo {pages} {114501} (\bibinfo {year} {2023})}\BibitemShut
  {NoStop}%
  \bibitem {Aoki}%
  \BibitemOpen
  \bibfield  {author} {\bibinfo {author} {\bibfnamefont {D}~\bibnamefont
  {Aoki}}, \bibinfo {author} {\bibfnamefont {J-P}~\bibnamefont
  {Brison}}, \bibinfo {author} {\bibfnamefont {J}~\bibnamefont
  {Flouquet}}, \bibinfo {author} {\bibfnamefont {K}~\bibnamefont
  {Ishida}}, \bibinfo {author} {\bibfnamefont {Y}~\bibnamefont
  {Tokunaga}}, and \bibinfo {author} {\bibfnamefont {Y}~\bibnamefont
  {Yanase}}\ }\bibfield  {title} {\enquote
  {\bibinfo {title} {{Unconventional superconductivity in UTe$_2$}},}\ }\href
  {https://doi.org/10.1088/1361-648X/ac5863} {\bibfield  {journal} {\bibinfo
  {journal} {Journal of Physics: Condensed Matter}\ }\textbf {\bibinfo {volume}
  {34}},\ \bibinfo {pages} {243002} (\bibinfo {year} {2022})}\BibitemShut
  {NoStop}%
 \bibitem {Joynt02}%
  \BibitemOpen
  \bibfield  {author} {\bibinfo {author} {\bibfnamefont {Robert}~\bibnamefont
  {Joynt}}, and \bibinfo {author} {\bibfnamefont {Louis}~\bibnamefont
  {Taillefer}} }\bibfield  {title} {\enquote
  {\bibinfo {title} {{The superconducting phase of UPt$_3$}},}\ }\href
  {https://doi.org/10.1103/RevModPhys.74.235} {\bibfield  {journal} {\bibinfo
  {journal} {Rev. Mod. Phys.}\ }\textbf {\bibinfo {volume}
  {74}},\ \bibinfo {pages} {235} (\bibinfo {year} {2002})}\BibitemShut
  {NoStop}%
  \bibitem {Bisset25}%
  \BibitemOpen
  \bibfield  {author} {\bibinfo {author} {\bibfnamefont {Rebecca}~\bibnamefont
  {Bisset}}, \bibinfo {author} {\bibfnamefont {Luke C}~\bibnamefont
  {Rhodes}}, \bibinfo {author} {\bibfnamefont {Hugo}~\bibnamefont
  {Decitre}}, \bibinfo {author} {\bibfnamefont {Matthew J}~\bibnamefont
  {Neat}}, \bibinfo {author} {\bibfnamefont {Ana}~\bibnamefont
  {Maldonado}}, \bibinfo {author} {\bibfnamefont {Andrew}~\bibnamefont
  {Huxley}}, \bibinfo {author} {\bibfnamefont {Carolina A}~\bibnamefont
  {Marques}}, and \bibinfo {author} {\bibfnamefont {P}~\bibnamefont
  {Wahl}}\ }\bibfield  {title} {\enquote
  {\bibinfo {title} {{Determining the superconducting order parameter of UPt$_3$ using scanning tunneling microscopy}},}\ }\href
  {https://doi.org//10.48550/arXiv.2512.15845} {\bibfield  {journal} {\bibinfo
  {journal} {Arxiv 2512.15845}\ }}\BibitemShut
  {NoStop}%
   \bibitem {Garcia}%
  \BibitemOpen
  \bibfield  {author} {\bibinfo {author} {\bibfnamefont {Pablo}~\bibnamefont
  {Garc\'ia~Talavera}}, and \bibinfo {author} {\bibfnamefont {et.}~\bibnamefont
  {al.}}\ }\bibfield  {title} {\enquote
  {\bibinfo {title} {{Surface charge density wave in UTe$_2$}},}\ }\href
  {https://doi.org//10.48550/arXiv.2504.12505} {\bibfield  {journal} {\bibinfo
  {journal} {Arxiv 2504.12505}\ }}\BibitemShut
  {NoStop}%
    \bibitem {Herrera}%
  \BibitemOpen
  \bibfield  {author} {\bibinfo {author} {\bibfnamefont {Edwin}~\bibnamefont
  {Herrera}}, and \bibinfo {author} {\bibfnamefont {et.}~\bibnamefont
  {al.}}\ }\bibfield  {title} {\enquote
  {\bibinfo {title} {{Quantum-well states at the surface of a heavy-fermion superconductor}},}\ }\href
  {https://doi.org//10.1038/s41586-023-05830-1} {\bibfield  {journal} {\bibinfo
  {journal} {Nature}\ }\textbf {\bibinfo {volume}
  {616}},\ \bibinfo {pages} {465} (\bibinfo {year} {2023})}\BibitemShut
  {NoStop}%
   \bibitem {Moreno}%
  \BibitemOpen
  \bibfield  {author} {\bibinfo {author} {\bibfnamefont {Jose~Antonio}~\bibnamefont
  {Moreno}}, and \bibinfo {author} {\bibfnamefont {et.}~\bibnamefont
  {al.}}\ }\bibfield  {title} {\enquote
  {\bibinfo {title} {{Quantum-well states at the surface of a heavy-fermion superconductor}},}\ }\href
  {https://doi.org/10.48550/arXiv.2508.04867} {\bibfield  {journal} {\bibinfo
  {journal} {Arxiv 2508.04867}\ } }\BibitemShut
  {NoStop}%
  \bibitem {Schneider}%
  \BibitemOpen
  \bibfield  {author} {\bibinfo {author} {\bibfnamefont {Lucas}~\bibnamefont
  {Schneider}}, \bibinfo {author} {\bibfnamefont {Khai That}~\bibnamefont
  {Ton}}, \bibinfo {author} {\bibfnamefont {Ioannis}~\bibnamefont
  {Ioannidis}}, \bibinfo {author} {\bibfnamefont {Jannis}~\bibnamefont
  {Neuhaus-Steinmetz}}, \bibinfo {author} {\bibfnamefont {Thore}~\bibnamefont
  {Posske}}, \bibinfo {author} {\bibfnamefont {Roland}~\bibnamefont
  {Wiesendanger}}, and \bibinfo {author} {\bibfnamefont {Jens}~\bibnamefont
  {Wiebe}}\ }\bibfield  {title} {\enquote
  {\bibinfo {title} {{Proximity superconductivity in atom-by-atom crafted quantum dots}},}\ }\href
  {https://doi.org/10.1038/s41586-023-06312-0} {\bibfield  {journal} {\bibinfo
  {journal} {Nature}\ }\textbf {\bibinfo {volume}
  {621}},\ \bibinfo {pages} {60} (\bibinfo {year} {2023})}\BibitemShut
  {NoStop}%
    \bibitem [{\citenamefont {Strigl}\ \emph {et~al.}(2015)\citenamefont {Strigl},
  \citenamefont {Espy}, \citenamefont {Bückle}, \citenamefont {Scheer},\ and\
  \citenamefont {Pietsch}}]{Strigle2015}%
  \BibitemOpen
  \bibfield  {author} {\bibinfo {author} {\bibfnamefont {F.}~\bibnamefont
  {Strigl}}, \bibinfo {author} {\bibfnamefont {C.}~\bibnamefont {Espy}},
  \bibinfo {author} {\bibfnamefont {M.}~\bibnamefont {Bückle}}, \bibinfo
  {author} {\bibfnamefont {E.}~\bibnamefont {Scheer}},\ and\ \bibinfo {author}
  {\bibfnamefont {T.}~\bibnamefont {Pietsch}},\ }\bibfield  {title} {\enquote
  {\bibinfo {title} {{Emerging magnetic order in platinum atomic contacts and
  chains}},}\ }\href {https://doi.org/10.1038/ncomms7172} {\bibfield  {journal}
  {\bibinfo  {journal} {Nature Communications}\ }\textbf {\bibinfo {volume}
  {6}},\ \bibinfo {pages} {6172} (\bibinfo {year} {2015})}\BibitemShut
  {NoStop}%
\bibitem [{\citenamefont {Rodrigues}\ \emph {et~al.}(2003)\citenamefont
  {Rodrigues}, \citenamefont {Bettini}, \citenamefont {Silva},\ and\
  \citenamefont {Ugarte}}]{Rodrigues2003Co}%
  \BibitemOpen
  \bibfield  {author} {\bibinfo {author} {\bibfnamefont {V.}~\bibnamefont
  {Rodrigues}}, \bibinfo {author} {\bibfnamefont {J.}~\bibnamefont {Bettini}},
  \bibinfo {author} {\bibfnamefont {P.~C.}\ \bibnamefont {Silva}},\ and\
  \bibinfo {author} {\bibfnamefont {D.}~\bibnamefont {Ugarte}},\ }\bibfield
  {title} {\enquote {\bibinfo {title} {Evidence for spontaneous spin-polarized
  transport in magnetic nanowires},}\ }\href
  {https://doi.org/10.1103/PhysRevLett.91.096801} {\bibfield  {journal}
  {\bibinfo  {journal} {Physical Review Letters}\ }\textbf {\bibinfo {volume}
  {91}},\ \bibinfo {pages} {096801} (\bibinfo {year} {2003})}\BibitemShut
  {NoStop}%
\bibitem [{\citenamefont {Untiedt}\ \emph {et~al.}(2004)\citenamefont
  {Untiedt}, \citenamefont {Dekker}, \citenamefont {Djukic},\ and\
  \citenamefont {van Ruitenbeek}}]{Untiedt2004}%
  \BibitemOpen
  \bibfield  {author} {\bibinfo {author} {\bibfnamefont {C.}~\bibnamefont
  {Untiedt}}, \bibinfo {author} {\bibfnamefont {D.~M.~T.}\ \bibnamefont
  {Dekker}}, \bibinfo {author} {\bibfnamefont {D.}~\bibnamefont {Djukic}},\
  and\ \bibinfo {author} {\bibfnamefont {J.~M.}\ \bibnamefont {van
  Ruitenbeek}},\ }\bibfield  {title} {\enquote {\bibinfo {title} {{Absence of
  magnetically induced fractional quantization in atomic contacts}},}\ }\href
  {https://doi.org/10.1103/PhysRevB.69.081401} {\bibfield  {journal} {\bibinfo
  {journal} {Physical Review B}\ }\textbf {\bibinfo {volume} {69}},\ \bibinfo
  {pages} {081401} (\bibinfo {year} {2004})}\BibitemShut {NoStop}%
\bibitem [{\citenamefont {Fern\'andez-Rossier}\ \emph
  {et~al.}(2005)\citenamefont {Fern\'andez-Rossier}, \citenamefont {Jacob},
  \citenamefont {Untiedt},\ and\ \citenamefont {Palacios}}]{Fernandez2005}%
  \BibitemOpen
  \bibfield  {author} {\bibinfo {author} {\bibfnamefont {J.}~\bibnamefont
  {Fern\'andez-Rossier}}, \bibinfo {author} {\bibfnamefont {D.}~\bibnamefont
  {Jacob}}, \bibinfo {author} {\bibfnamefont {C.}~\bibnamefont {Untiedt}},\
  and\ \bibinfo {author} {\bibfnamefont {J.~J.}\ \bibnamefont {Palacios}},\
  }\bibfield  {title} {\enquote {\bibinfo {title} {{Transport in magnetically
  ordered Pt nanocontacts}},}\ }\href
  {https://doi.org/10.1103/PhysRevB.72.224418} {\bibfield  {journal} {\bibinfo
  {journal} {Physical Review B}\ }\textbf {\bibinfo {volume} {72}},\ \bibinfo
  {pages} {224418} (\bibinfo {year} {2005})}\BibitemShut {NoStop}%
\bibitem [{\citenamefont {Kinikar}\ \emph {et~al.}(2017)\citenamefont
  {Kinikar}, \citenamefont {Phanindra~Sai}, \citenamefont {Bhattacharyya},
  \citenamefont {Agarwala}, \citenamefont {Biswas}, \citenamefont {Sarker},
  \citenamefont {Krishnamurthy}, \citenamefont {Jain}, \citenamefont {Shenoy},\
  and\ \citenamefont {Ghosh}}]{Kinikar2017}%
  \BibitemOpen
  \bibfield  {author} {\bibinfo {author} {\bibfnamefont {A.}~\bibnamefont
  {Kinikar}}, \bibinfo {author} {\bibfnamefont {T.}~\bibnamefont
  {Phanindra~Sai}}, \bibinfo {author} {\bibfnamefont {S.}~\bibnamefont
  {Bhattacharyya}}, \bibinfo {author} {\bibfnamefont {A.}~\bibnamefont
  {Agarwala}}, \bibinfo {author} {\bibfnamefont {T.}~\bibnamefont {Biswas}},
  \bibinfo {author} {\bibfnamefont {S.~K.}\ \bibnamefont {Sarker}}, \bibinfo
  {author} {\bibfnamefont {H.~R.}\ \bibnamefont {Krishnamurthy}}, \bibinfo
  {author} {\bibfnamefont {M.}~\bibnamefont {Jain}}, \bibinfo {author}
  {\bibfnamefont {V.~B.}\ \bibnamefont {Shenoy}},\ and\ \bibinfo {author}
  {\bibfnamefont {A.}~\bibnamefont {Ghosh}},\ }\bibfield  {title} {\enquote
  {\bibinfo {title} {{Quantized edge modes in atomic-scale point contacts in
  graphene}},}\ }\href {https://doi.org/10.1038/nnano.2017.24} {\bibfield
  {journal} {\bibinfo  {journal} {Nature Nanotechnology}\ }\textbf {\bibinfo
  {volume} {12}},\ \bibinfo {pages} {564--568} (\bibinfo {year}
  {2017})}\BibitemShut {NoStop}%
\bibitem [{\citenamefont {Kawamura}\ \emph {et~al.}(2015)\citenamefont
  {Kawamura}, \citenamefont {Ono}, \citenamefont {Stano}, \citenamefont
  {Kono},\ and\ \citenamefont {Aono}}]{PhysRevLett.115.036601}%
  \BibitemOpen
  \bibfield  {author} {\bibinfo {author} {\bibfnamefont {M.}~\bibnamefont
  {Kawamura}}, \bibinfo {author} {\bibfnamefont {K.}~\bibnamefont {Ono}},
  \bibinfo {author} {\bibfnamefont {P.}~\bibnamefont {Stano}}, \bibinfo
  {author} {\bibfnamefont {K.}~\bibnamefont {Kono}},\ and\ \bibinfo {author}
  {\bibfnamefont {T.}~\bibnamefont {Aono}},\ }\bibfield  {title} {\enquote
  {\bibinfo {title} {{Electronic magnetization of a quantum point contact
  measured by nuclear magnetic resonance}},}\ }\href
  {https://doi.org/10.1103/PhysRevLett.115.036601} {\bibfield  {journal}
  {\bibinfo  {journal} {Physical Review Letters}\ }\textbf {\bibinfo {volume}
  {115}},\ \bibinfo {pages} {036601} (\bibinfo {year} {2015})}\BibitemShut
  {NoStop}%
\bibitem [{\citenamefont {Brun}\ \emph {et~al.}(2014)\citenamefont {Brun},
  \citenamefont {Martins}, \citenamefont {Faniel}, \citenamefont {Hackens},
  \citenamefont {Bachelier}, \citenamefont {Cavanna}, \citenamefont {Ulysse},
  \citenamefont {Ouerghi}, \citenamefont {Gennser}, \citenamefont {Mailly},
  \citenamefont {Huant}, \citenamefont {Bayot}, \citenamefont {Sanquer},\ and\
  \citenamefont {Sellier}}]{Brun2014}%
  \BibitemOpen
  \bibfield  {author} {\bibinfo {author} {\bibfnamefont {B.}~\bibnamefont
  {Brun}}, \bibinfo {author} {\bibfnamefont {F.}~\bibnamefont {Martins}},
  \bibinfo {author} {\bibfnamefont {S.}~\bibnamefont {Faniel}}, \bibinfo
  {author} {\bibfnamefont {B.}~\bibnamefont {Hackens}}, \bibinfo {author}
  {\bibfnamefont {G.}~\bibnamefont {Bachelier}}, \bibinfo {author}
  {\bibfnamefont {A.}~\bibnamefont {Cavanna}}, \bibinfo {author} {\bibfnamefont
  {C.}~\bibnamefont {Ulysse}}, \bibinfo {author} {\bibfnamefont
  {A.}~\bibnamefont {Ouerghi}}, \bibinfo {author} {\bibfnamefont
  {U.}~\bibnamefont {Gennser}}, \bibinfo {author} {\bibfnamefont
  {D.}~\bibnamefont {Mailly}}, \bibinfo {author} {\bibfnamefont
  {et~al.}},\ }\bibfield  {title} {\enquote {\bibinfo
  {title} {{Wigner and Kondo physics in quantum point contacts revealed by
  scanning gate microscopy}},}\ }\href {https://doi.org/10.1038/ncomms5290}
  {\bibfield  {journal} {\bibinfo  {journal} {Nature Communications}\ }\textbf
  {\bibinfo {volume} {5}},\ \bibinfo {pages} {4290} (\bibinfo {year}
  {2014})}\BibitemShut {NoStop}%
\bibitem [{\citenamefont {Bagrets}, \citenamefont {Papanikolaou},\ and\
  \citenamefont {Mertig}(2004)}]{PhysRevB.70.064410}%
  \BibitemOpen
  \bibfield  {author} {\bibinfo {author} {\bibfnamefont {A.}~\bibnamefont
  {Bagrets}}, \bibinfo {author} {\bibfnamefont {N.}~\bibnamefont
  {Papanikolaou}},\ and\ \bibinfo {author} {\bibfnamefont {I.}~\bibnamefont
  {Mertig}},\ }\bibfield  {title} {\enquote {\bibinfo {title}
  {{Magnetoresistance of atomic-sized contacts: An ab initio study}},}\ }\href
  {https://doi.org/10.1103/PhysRevB.70.064410} {\bibfield  {journal} {\bibinfo
  {journal} {Physical Review B}\ }\textbf {\bibinfo {volume} {70}},\ \bibinfo
  {pages} {064410} (\bibinfo {year} {2004})}\BibitemShut {NoStop}%
\bibitem [{\citenamefont {Jacob}, \citenamefont {Fern\'andez-Rossier},\ and\
  \citenamefont {Palacios}(2005)}]{PhysRevB.71.220403}%
  \BibitemOpen
  \bibfield  {author} {\bibinfo {author} {\bibfnamefont {D.}~\bibnamefont
  {Jacob}}, \bibinfo {author} {\bibfnamefont {J.}~\bibnamefont
  {Fern\'andez-Rossier}},\ and\ \bibinfo {author} {\bibfnamefont {J.~J.}\
  \bibnamefont {Palacios}},\ }\bibfield  {title} {\enquote {\bibinfo {title}
  {{Magnetic and orbital blocking in Ni nanocontacts}},}\ }\href
  {https://doi.org/10.1103/PhysRevB.71.220403} {\bibfield  {journal} {\bibinfo
  {journal} {Physical Review B}\ }\textbf {\bibinfo {volume} {71}},\ \bibinfo
  {pages} {220403} (\bibinfo {year} {2005})}\BibitemShut {NoStop}%
\bibitem [{\citenamefont {Suderow}\ \emph {et~al.}(2003)\citenamefont
  {Suderow}, \citenamefont {Crespo}, \citenamefont {Vieira}, \citenamefont
  {Vila}, \citenamefont {Garc\'ia-Hern\'andez}, \citenamefont {{de Andr\'es}},
  \citenamefont {Prieto}, \citenamefont {Ocal}, \citenamefont {Mart\'inez},\
  and\ \citenamefont {Mukovskii}}]{SUDEROW2003264}%
  \BibitemOpen
  \bibfield  {author} {\bibinfo {author} {\bibfnamefont {H.}~\bibnamefont
  {Suderow}}, \bibinfo {author} {\bibfnamefont {M.}~\bibnamefont {Crespo}},
  \bibinfo {author} {\bibfnamefont {S.}~\bibnamefont {Vieira}}, \bibinfo
  {author} {\bibfnamefont {M.}~\bibnamefont {Vila}}, \bibinfo {author}
  {\bibfnamefont {M.}~\bibnamefont {Garc\'ia-Hern\'andez}}, \bibinfo {author}
  {\bibfnamefont {A.}~\bibnamefont {{de Andr\'es}}}, \bibinfo {author}
  {\bibfnamefont {C.}~\bibnamefont {Prieto}}, \bibinfo {author} {\bibfnamefont
  {C.}~\bibnamefont {Ocal}}, \bibinfo {author} {\bibfnamefont {J.}~\bibnamefont
  {Mart\'inez}},\ and\ \bibinfo {author} {\bibfnamefont {Y.}~\bibnamefont
  {Mukovskii}},\ }\bibfield  {title} {\enquote {\bibinfo {title} {{Observation
  of a spin-polarized current through single atom quantum point contacts}},}\
  }\href {https://doi.org/10.1016/S1386-9477(02)01010-X} {\bibfield  {journal}
  {\bibinfo  {journal} {Physica E: Low-dimensional Systems and Nanostructures}\
  }\textbf {\bibinfo {volume} {18}},\ \bibinfo {pages} {264--265} (\bibinfo
  {year} {2003})},\ \bibinfo {note} {23rd International Conference on Low
  Temperature Physics (LT23)}\BibitemShut {NoStop}%
\bibitem [{\citenamefont {Doudin}\ and\ \citenamefont
  {Viret}(2008)}]{Doudin2008}%
  \BibitemOpen
  \bibfield  {author} {\bibinfo {author} {\bibfnamefont {B.}~\bibnamefont
  {Doudin}}\ and\ \bibinfo {author} {\bibfnamefont {M.}~\bibnamefont {Viret}},\
  }\bibfield  {title} {\enquote {\bibinfo {title} {{Ballistic
  magnetoresistance?}}}\ }\href {https://doi.org/10.1088/0953-8984/20/8/083201}
  {\bibfield  {journal} {\bibinfo  {journal} {Journal of Physics: Condensed
  Matter}\ }\textbf {\bibinfo {volume} {20}},\ \bibinfo {pages} {083201}
  (\bibinfo {year} {2008})}\BibitemShut {NoStop}%
\bibitem [{\citenamefont {Calvo}\ \emph {et~al.}(2009)\citenamefont {Calvo},
  \citenamefont {Fern\'andez-Rossier}, \citenamefont {Palacios}, \citenamefont
  {Jacob}, \citenamefont {Natelson},\ and\ \citenamefont
  {Untiedt}}]{Calvo2009}%
  \BibitemOpen
  \bibfield  {author} {\bibinfo {author} {\bibfnamefont {M.~R.}\ \bibnamefont
  {Calvo}}, \bibinfo {author} {\bibfnamefont {J.}~\bibnamefont
  {Fern\'andez-Rossier}}, \bibinfo {author} {\bibfnamefont {J.~J.}\
  \bibnamefont {Palacios}}, \bibinfo {author} {\bibfnamefont {D.}~\bibnamefont
  {Jacob}}, \bibinfo {author} {\bibfnamefont {D.}~\bibnamefont {Natelson}},\
  and\ \bibinfo {author} {\bibfnamefont {C.}~\bibnamefont {Untiedt}},\
  }\bibfield  {title} {\enquote {\bibinfo {title} {{The Kondo effect in
  ferromagnetic atomic contacts}},}\ }\href
  {https://doi.org/10.1038/nature07878} {\bibfield  {journal} {\bibinfo
  {journal} {Nature}\ }\textbf {\bibinfo {volume} {458}},\ \bibinfo {pages}
  {1150--1153} (\bibinfo {year} {2009})}\BibitemShut {NoStop}%
\bibitem [{\citenamefont {Vardimon}, \citenamefont {Klionsky},\ and\
  \citenamefont {Tal}(2015)}]{Vardimon2015}%
  \BibitemOpen
  \bibfield  {author} {\bibinfo {author} {\bibfnamefont {R.}~\bibnamefont
  {Vardimon}}, \bibinfo {author} {\bibfnamefont {M.}~\bibnamefont {Klionsky}},\
  and\ \bibinfo {author} {\bibfnamefont {O.}~\bibnamefont {Tal}},\ }\bibfield
  {title} {\enquote {\bibinfo {title} {Indication of complete spin filtering in
  atomic-scale nickel oxide},}\ }\href
  {https://doi.org/10.1021/acs.nanolett.5b00729} {\bibfield  {journal}
  {\bibinfo  {journal} {Nano Letters}\ }\textbf {\bibinfo {volume} {15}},\
  \bibinfo {pages} {3894--3898} (\bibinfo {year} {2015})}\BibitemShut {NoStop}%
\bibitem [{\citenamefont {Rakhmilevitch}\ \emph {et~al.}(2016)\citenamefont
  {Rakhmilevitch}, \citenamefont {Sarkar}, \citenamefont {Bitton},
  \citenamefont {Kronik},\ and\ \citenamefont {Tal}}]{Rakhmilevitch2016}%
  \BibitemOpen
  \bibfield  {author} {\bibinfo {author} {\bibfnamefont {D.}~\bibnamefont
  {Rakhmilevitch}}, \bibinfo {author} {\bibfnamefont {S.}~\bibnamefont
  {Sarkar}}, \bibinfo {author} {\bibfnamefont {O.}~\bibnamefont {Bitton}},
  \bibinfo {author} {\bibfnamefont {L.}~\bibnamefont {Kronik}},\ and\ \bibinfo
  {author} {\bibfnamefont {O.}~\bibnamefont {Tal}},\ }\bibfield  {title}
  {\enquote {\bibinfo {title} {Enhanced magnetoresistance in molecular
  junctions by geometrical optimization of spin-selective orbital
  hybridization},}\ }\href {https://doi.org/10.1021/acs.nanolett.5b04674}
  {\bibfield  {journal} {\bibinfo  {journal} {Nano Letters}\ }\textbf {\bibinfo
  {volume} {16}},\ \bibinfo {pages} {1741--1745} (\bibinfo {year}
  {2016})}\BibitemShut {NoStop}%
\bibitem [{\citenamefont {Vardimon}\ \emph {et~al.}(2016)\citenamefont
  {Vardimon}, \citenamefont {Matt}, \citenamefont {Nielaba}, \citenamefont
  {Cuevas},\ and\ \citenamefont {Tal}}]{PhysRevB.93.085439}%
  \BibitemOpen
  \bibfield  {author} {\bibinfo {author} {\bibfnamefont {R.}~\bibnamefont
  {Vardimon}}, \bibinfo {author} {\bibfnamefont {M.}~\bibnamefont {Matt}},
  \bibinfo {author} {\bibfnamefont {P.}~\bibnamefont {Nielaba}}, \bibinfo
  {author} {\bibfnamefont {J.~C.}\ \bibnamefont {Cuevas}},\ and\ \bibinfo
  {author} {\bibfnamefont {O.}~\bibnamefont {Tal}},\ }\bibfield  {title}
  {\enquote {\bibinfo {title} {Orbital origin of the electrical conduction in
  ferromagnetic atomic-size contacts: Insights from shot noise measurements and
  theoretical simulations},}\ }\href
  {https://doi.org/10.1103/PhysRevB.93.085439} {\bibfield  {journal} {\bibinfo
  {journal} {Phys. Rev. B}\ }\textbf {\bibinfo {volume} {93}},\ \bibinfo
  {pages} {085439} (\bibinfo {year} {2016})}\BibitemShut {NoStop}%
\bibitem [{\citenamefont {Chakrabarti}\ \emph {et~al.}(2022)\citenamefont
  {Chakrabarti}, \citenamefont {Vilan}, \citenamefont {Deutch}, \citenamefont
  {Oz}, \citenamefont {Hod}, \citenamefont {Peralta},\ and\ \citenamefont
  {Tal}}]{Chakrabarti2022}%
  \BibitemOpen
  \bibfield  {author} {\bibinfo {author} {\bibfnamefont {S.}~\bibnamefont
  {Chakrabarti}}, \bibinfo {author} {\bibfnamefont {A.}~\bibnamefont {Vilan}},
  \bibinfo {author} {\bibfnamefont {G.}~\bibnamefont {Deutch}}, \bibinfo
  {author} {\bibfnamefont {A.}~\bibnamefont {Oz}}, \bibinfo {author}
  {\bibfnamefont {O.}~\bibnamefont {Hod}}, \bibinfo {author} {\bibfnamefont
  {J.~E.}\ \bibnamefont {Peralta}},\ and\ \bibinfo {author} {\bibfnamefont
  {O.}~\bibnamefont {Tal}},\ }\bibfield  {title} {\enquote {\bibinfo {title}
  {Magnetic control over the fundamental structure of atomic wires},}\ }\href
  {https://doi.org/10.1038/s41467-022-31456-4} {\bibfield  {journal} {\bibinfo
  {journal} {Nature Communications}\ }\textbf {\bibinfo {volume} {13}},\
  \bibinfo {pages} {4113} (\bibinfo {year} {2022})}\BibitemShut {NoStop}%
\bibitem [{\citenamefont {Hayakawa}\ \emph {et~al.}(2016)\citenamefont
  {Hayakawa}, \citenamefont {Karimi}, \citenamefont {Wolf}, \citenamefont
  {Huhn}, \citenamefont {Z{\"o}llner}, \citenamefont {Herrmann},\ and\
  \citenamefont {Scheer}}]{Hayakawa2016}%
  \BibitemOpen
  \bibfield  {author} {\bibinfo {author} {\bibfnamefont {R.}~\bibnamefont
  {Hayakawa}}, \bibinfo {author} {\bibfnamefont {M.~A.}\ \bibnamefont
  {Karimi}}, \bibinfo {author} {\bibfnamefont {J.}~\bibnamefont {Wolf}},
  \bibinfo {author} {\bibfnamefont {T.}~\bibnamefont {Huhn}}, \bibinfo {author}
  {\bibfnamefont {M.~S.}\ \bibnamefont {Z{\"o}llner}}, \bibinfo {author}
  {\bibfnamefont {C.}~\bibnamefont {Herrmann}},\ and\ \bibinfo {author}
  {\bibfnamefont {E.}~\bibnamefont {Scheer}},\ }\bibfield  {title} {\enquote
  {\bibinfo {title} {Large magnetoresistance in single-radical molecular
  junctions},}\ }\href {https://doi.org/10.1021/acs.nanolett.6b01595}
  {\bibfield  {journal} {\bibinfo  {journal} {Nano Letters}\ }\textbf {\bibinfo
  {volume} {16}},\ \bibinfo {pages} {4960--4967} (\bibinfo {year}
  {2016})}\BibitemShut {NoStop}%
\bibitem [{\citenamefont {Egle}\ \emph {et~al.}(2010)\citenamefont {Egle},
  \citenamefont {Bacca}, \citenamefont {Pernau}, \citenamefont {Huefner},
  \citenamefont {Hinzke}, \citenamefont {Nowak},\ and\ \citenamefont
  {Scheer}}]{PhysRevB.81.134402}%
  \BibitemOpen
  \bibfield  {author} {\bibinfo {author} {\bibfnamefont {S.}~\bibnamefont
  {Egle}}, \bibinfo {author} {\bibfnamefont {C.}~\bibnamefont {Bacca}},
  \bibinfo {author} {\bibfnamefont {H.-F.}\ \bibnamefont {Pernau}}, \bibinfo
  {author} {\bibfnamefont {M.}~\bibnamefont {Huefner}}, \bibinfo {author}
  {\bibfnamefont {D.}~\bibnamefont {Hinzke}}, \bibinfo {author} {\bibfnamefont
  {U.}~\bibnamefont {Nowak}},\ and\ \bibinfo {author} {\bibfnamefont
  {E.}~\bibnamefont {Scheer}},\ }\bibfield  {title} {\enquote {\bibinfo {title}
  {Magnetoresistance of atomic-size contacts realized with mechanically
  controllable break junctions},}\ }\href
  {https://doi.org/10.1103/PhysRevB.81.134402} {\bibfield  {journal} {\bibinfo
  {journal} {Phys. Rev. B}\ }\textbf {\bibinfo {volume} {81}},\ \bibinfo
  {pages} {134402} (\bibinfo {year} {2010})}\BibitemShut {NoStop}%
\bibitem [{\citenamefont {Strigl}\ \emph {et~al.}(2016)\citenamefont {Strigl},
  \citenamefont {Keller}, \citenamefont {Weber}, \citenamefont {Pietsch},\ and\
  \citenamefont {Scheer}}]{PhysRevB.94.144431}%
  \BibitemOpen
  \bibfield  {author} {\bibinfo {author} {\bibfnamefont {F.}~\bibnamefont
  {Strigl}}, \bibinfo {author} {\bibfnamefont {M.}~\bibnamefont {Keller}},
  \bibinfo {author} {\bibfnamefont {D.}~\bibnamefont {Weber}}, \bibinfo
  {author} {\bibfnamefont {T.}~\bibnamefont {Pietsch}},\ and\ \bibinfo {author}
  {\bibfnamefont {E.}~\bibnamefont {Scheer}},\ }\bibfield  {title} {\enquote
  {\bibinfo {title} {{Magnetism in Pd: Magnetoconductance and transport
  spectroscopy of atomic contacts}},}\ }\href
  {https://doi.org/10.1103/PhysRevB.94.144431} {\bibfield  {journal} {\bibinfo
  {journal} {Phys. Rev. B}\ }\textbf {\bibinfo {volume} {94}},\ \bibinfo
  {pages} {144431} (\bibinfo {year} {2016})}\BibitemShut {NoStop}%
\bibitem [{\citenamefont {Prestel}\ \emph {et~al.}(2019)\citenamefont
  {Prestel}, \citenamefont {Ritter}, \citenamefont {Di~Bernardo}, \citenamefont
  {Pietsch},\ and\ \citenamefont {Scheer}}]{PhysRevB.100.214439}%
  \BibitemOpen
  \bibfield  {author} {\bibinfo {author} {\bibfnamefont {M.~W.}\ \bibnamefont
  {Prestel}}, \bibinfo {author} {\bibfnamefont {M.~F.}\ \bibnamefont {Ritter}},
  \bibinfo {author} {\bibfnamefont {A.}~\bibnamefont {Di~Bernardo}}, \bibinfo
  {author} {\bibfnamefont {T.}~\bibnamefont {Pietsch}},\ and\ \bibinfo {author}
  {\bibfnamefont {E.}~\bibnamefont {Scheer}},\ }\bibfield  {title} {\enquote
  {\bibinfo {title} {Tuning the magnetic anisotropy energy of atomic wires},}\
  }\href {https://doi.org/10.1103/PhysRevB.100.214439} {\bibfield  {journal}
  {\bibinfo  {journal} {Phys. Rev. B}\ }\textbf {\bibinfo {volume} {100}},\
  \bibinfo {pages} {214439} (\bibinfo {year} {2019})}\BibitemShut {NoStop}%
\bibitem [{\citenamefont {Wu}\ \emph {et~al.}(2024)\citenamefont {Wu},
  \citenamefont {Martínez}, \citenamefont {Obladen}, \citenamefont
  {Fernández-Lomana}, \citenamefont {Herrera}, \citenamefont {Sabater},
  \citenamefont {Palacios}, \citenamefont {Guillamón},\ and\ \citenamefont
  {Suderow}}]{wu2024magnetic}%
  \BibitemOpen
  \bibfield  {author} {\bibinfo {author} {\bibfnamefont {B.}~\bibnamefont
  {Wu}}, \bibinfo {author} {\bibfnamefont {A.}~\bibnamefont {Martínez}},
  \bibinfo {author} {\bibfnamefont {P.}~\bibnamefont {Obladen}}, \bibinfo
  {author} {\bibfnamefont {M.}~\bibnamefont {Fernández-Lomana}}, \bibinfo
  {author} {\bibfnamefont {E.}~\bibnamefont {Herrera}}, \bibinfo {author}
  {\bibfnamefont {C.}~\bibnamefont {Sabater}}, \bibinfo {author} {\bibfnamefont
  {J.~J.}\ \bibnamefont {Palacios}}, \bibinfo {author} {\bibfnamefont
  {I.}~\bibnamefont {Guillamón}},\ and\ \bibinfo {author} {\bibfnamefont
  {H.}~\bibnamefont {Suderow}},\ }\href@noop {} {\enquote {\bibinfo {title}
  {Magnetic field dependence of the atomic and electronic structure of
  monovalent metallic nanocontacts unveiled in transport experiments},}\ }
  (\bibinfo {year} {2024}),\ \Eprint {https://arxiv.org/abs/2407.20577}
  {arXiv:2407.20577 [cond-mat.mes-hall]} \BibitemShut {NoStop}%
\end{thebibliography}
%aipnum4-2.bst 2019-01-14 (MD) hand-edited version of apsrev4-1.bst

\end{document}